\documentclass[11pt, a4]{article}

\usepackage{stolenstyle}

\usepackage{amsmath,epsf,amssymb,latexsym,amsthm,setspace,bbm,array,pifont,enumerate}

\DeclareFontFamily{U}{rsf}{}
\DeclareFontShape{U}{rsf}{m}{n}{
  <5> <6> rsfs5 <7> <8> <9> rsfs7 <10-> rsfs10}{}
\DeclareMathAlphabet\Scr{U}{rsf}{m}{n}

\def\iden{{\mathbbm 1}}

\def\rep#1{{{\boldsymbol{#1}}}}
\def\brep#1{{{\overline{\boldsymbol{#1}}}}}

\def\C{{\mathbb C}}

\def\R{{\mathbb R}}
\def\Z{{\mathbb Z}}

\def\Gr{\operatorname{Gr}}

\def\Hom{\operatorname{Hom}}

\def\Tr{\operatorname{Tr}}

\def\im{\operatorname{im}}

\def\SO{\operatorname{SO}}
\def\SL{\operatorname{SL}}
\def\GL{\operatorname{GL}}

\def\GO{\operatorname{O{}}}
\def\SU{\operatorname{SU}}
\def\GU{\operatorname{U{}}}

\def\Spin{\operatorname{Spin}}

\def\GE{\operatorname{E}}

\def\so{\operatorname{\mathfrak{so}}}

\def\su{\operatorname{\mathfrak{su}}}
\def\Lu{\operatorname{\mathfrak{u}}}
\def\Le{\operatorname{\mathfrak{e}}}
\def\Lg{\operatorname{\mathfrak{g}}}

\def\p{\partial}
\def\pb{\bar{\partial}}

\def\ra{\rangle}

\def\ff#1#2{{\textstyle\frac{#1}{#2}}}

\def\cA{{\cal A}}

\def\cF{{\cal F}}
\def\cG{{\cal G}}
\def\cH{{\cal H}}

\def\cL{{\cal L}}

\def\cV{{\cal V}}

\def\cX{{\cal X}}

\def\cZ{{\cal Z}}

\def\ep{{\epsilon}}

\newcommand\gammab{\overline{\gamma}}

\newcommand\etab{\overline{\eta}}

\newcommand\xib{\overline{\xi}}

\newcommand\taub{\overline{\tau}}

\newcommand\psib{\overline{\psi}}

\newcommand\gammat{\widetilde{\gamma}}

\newcommand\pit{\widetilde{\pi}}
\newcommand\rhot{\widetilde{\rho}}

\newcommand\taut{\widetilde{\tau}}

\newcommand\cb{\overline{c}}

\newcommand\hb{\overline{h}}

\newcommand\pbar{\overline{p}} 
\newcommand\qb{\overline{q}}

\newcommand\wb{\overline{w}}

\newcommand\zb{\overline{z}}

\newcommand\htld{\widetilde{h}} 

\newcommand\nt{\widetilde{n}}

\newcommand\Hh{\widehat{H}}

\newcommand\Jb{\overline{J}}

\newcommand\Lb{\overline{L}}

\newcommand\Mt{\widetilde{M}}

\newcommand\Rt{\widetilde{R}}

\theoremstyle{definition}

\usepackage{tikz}

\usetikzlibrary[shapes.geometric]
\usetikzlibrary{positioning} 
\usetikzlibrary{calc,intersections,through,backgrounds}
\usetikzlibrary{cd}

\tikzset{>=stealth}
\tikzset{every picture/.style={very thick}}

\def\bqb{{{{\qb}}}}

\def\Ttor{{{\mathbb{T}}}}

\def\balpha{\boldsymbol{\alpha}}
\def\bC{\boldsymbol{C}}
\def\be{\boldsymbol{e}}
\def\bgamma{\boldsymbol{\gamma}}
\def\bpi{\boldsymbol{\pi}}
\def\bpit{\widetilde{\boldsymbol{\pi}}}

\def\bv{\boldsymbol{v}}

\def\bq{{\textbf{q}}}
\def\bqbar{{\overline{\bq}}}

\def\preprint{ ~~ }

\title{Flat equivariant gerbes: holonomies and dualities}
\author[a] {Peng Cheng,}
\author[b] {Ilarion V.~Melnikov,}
\author[a] {Ruben Minasian}
\affiliation[a] {Institut de Physique Th{\'e}orique,  
Universit{\'e} Paris-Saclay, CNRS, CEA, F-9119, Gif-sur-Yvette, France
}
\affiliation[b] {Department of Physics and Astronomy,
James Madison University,
Harrisonburg, VA 22807, USA}

\emailAdd{peng.cheng@ipht.fr}
\emailAdd{melnikix@jmu.edu}
\emailAdd{ruben.minasian@ipht.fr}

\abstract{We examine the role of global topological data associated to choices of holonomy for flat gauge fields in string compactification.  Our study begins with perturbative string compactification on compact flat manifolds preserving $8$ supercharges in $5$ dimensions.  By including non-trivial holonomy for Wilson lines in the heterotic string and for the B-field gerbe in the type II string we find worldsheet dualities that relate these backgrounds to other string compactifications.  While our simple examples allow for explicit analysis, the concepts and some of the methods extend to a broader class of compactifications and have implications for string dualities, perturbative and otherwise.
}

\begin{document}

\maketitle

\section{Introduction} \label{s:Introduction}

Anyone familiar with the cover of~\cite{Deligne:1999qp} knows that interactions between mathematicians and physicists involve an aspect of translation: while the former starts with global data that may have some local presentation in a set of coordinates or representatives, the latter happily constructs quantities from a combination of greek and latin letters, large and small.  Yet the two have much to say to each other, and over the years the language has remarkably converged.  Now physicists are happy to write papers on differential cohomology, and mathematicians write texts on renormalization in quantum field theory.  

In this work we will be mostly concerned with structures where the usual physics methodology is of no avail, and the gauge fields that we might typically write will all be zero.  Nevertheless, the global topological data will have a profound effect on the physics.  The examples that we study are particularly simple:  toroidal orbifold conformal theories and the associated string compactifications.

Much of the fascination with orbifold compactification on $Y/G$ comes from the presence of singular points, which correspond to points in $Y$ fixed by the action of $G$~\cite{Dixon:1985jw,Dixon:1986jc}.  Despite these singularities, the string theory on $Y/G$ can be well-behaved, and computations of various quantities from the worldsheet perspective have lead to important insights into geometry, such as the notion of stringy Hodge numbers and finer cohomological structures on orbifolds and algebraic varieties (e.g. see~\cite{MR2359514} for an introduction).  

The study of orbifolds has also led to insights into the moduli space of string compactifications, for example by showing that certain limits of compactification on smooth manifolds such as K3 or a Calabi-Yau $3$-fold where the geometry becomes singular can nevertheless be understood entirely in perturbative string theory.  In addition, orbifolds have provided insights into non-trivial dualities such as mirror symmetry.  They have also been used extensively in string phenomenology, and there are impressive classification results, including the description of all symmetric orbifolds of the six-torus $\Ttor^6$ that preserve spacetime supersymmetry~\cite{Fischer:2012qj}. 

In recent years orbifolds of toroidal compactifications have received renewed interest in relation to the swampland program; see, for example,~\cite{Fraiman:2018ebo,Font:2020rsk,Font:2021uyw,Cvetic:2021sjm,Fraiman:2021soq,Fraiman:2021hma}.  This recent work involves asymmetric orbifolds that act on the toroidal geometry by shifts, while having a non-trivial action on the gauge sector.  The results provide insights into the structure of disconnected components in the moduli space of toroidal compactification that generalize the CHL construction~\cite{Chaudhuri:1995fk,Chaudhuri:1995bf} and are intimately related to the study of heterotic compactifications with non-trivial flat connections~\cite{Lerche:1997rr,deBoer:2001wca} and related type II/M-theory/F-theory constructions with background RR fluxes and frozen singularities ~\cite{Kachru:1997bz,Tachikawa:2015wka}.  It will be useful to extend these efforts to cases with less spacetime supersymmetry, where classification will be even more challenging, but also the physics will be richer.  Our work is just a small indication of some of the structures that emerge in this larger setting.

In this paper we begin our study with $5$-dimensional compactifications on $X_5 = \Ttor^5/G$ that preserve half of spacetime supersymmetry, and where $G= \Z_N$ acts freely on $\Ttor^5$.\footnote{As the recent work~\cite{Acharya:2022shu} indicates, if one is willing to abandon spacetime supersymmetry, then compact flat manifolds already become interesting in compactifications to $7$ dimensions.} To obtain a free and supersymmetric action, we write $\Ttor^5 = \Ttor^4 \times S^1$, and choose $G$ to have a supersymmetric action on $\Ttor^4$, so that $\Ttor^4/G$ has holonomy $\Z_N \subset \SU(2)$. The orbifold $X_5$ is nevertheless smooth because the action on $\Ttor^4$ is combined with an order $N$ shift on the $5$-th circle.

Since these quotients lack fixed points, we can usefully discuss their physics from both supergravity and worldsheet perspectives.  For example, we can interpret these compactifications from a six-dimensional spacetime point of view:  it is possible to think of the theories as $S^1$ compactifications of a $6$-dimensional theory, where a holonomy is turned on for a discrete $\Z_N$ gauge symmetry; the holonomy breaks half of the supersymmetry and modifies the spectrum in other ways.  This point of view should be useful in fitting these compactifications in the framework of F-theory/type II --- heterotic duality, for example along the lines explored in~\cite{Hull:2017llx,Gautier:2019qiq}.  

Second, we show that in heterotic string theory these compactifications are dual to more familiar compactifications on $\Ttor^4/G \times S^1$ and are therefore connected to conventional compactification on K3.  That this should be the case was already suggested in~\cite{Kachru:1995wm}, and we establish the claim at the level of worldsheet CFT.  In particular, we show that a small radius limit of the $X_5$ compactification develops additional massless states and a Higgs branch invisible in the supergravity description.  We also note that the duality exchanges spaces with different topologies, thus providing another example of stringy geometry relating topologically distinct geometries.

The equivalence is based on the observation that there are orbifold actions on the $\Ttor^5$ worldsheet CFT, $G_1\simeq G$ and $G_2 \simeq G$ that are related by a conjugation:  there is an element $t \in \GO(21,5,\Z)$ that gives an isomorphism $t: G_1 \to G_2$, with $t:  g \mapsto t^{-1} g t$.  While the quotient by $G_1$ leads to the smooth geometry $X_5$, the quotient by $G_2$ leads to $\Ttor^4/G \times S^1$.  The induced isomorphism on the moduli of the two CFTs acts as a T-duality on the radius of the non-trivial $S^1$ in combination with a shift of the Wilson line parameters by a lattice vector.  

It is then natural to interpret the isomorphism as a T-duality of the $X_5$ CFT, but there is a subtlety in applying the standard paradigm of Buscher rules for a non-linear sigma model, despite the fact that the $X_5$ CFT can be perfectly well described by a large radius non-linear sigma model.  The trouble is that the natural circle fibration structure
\begin{equation*}
\begin{tikzcd}
S^1 \arrow[r,hook] & X_5 \arrow[d] \\ & M
\end{tikzcd}
\end{equation*}
has $M = \Ttor^4/G$, i.e. the base is singular!  Our results indicate that T-duality extends to such singular fibrations, with the benefit of enlarging the possibilities for topology change via T-duality beyond those discussed in the context of principal torus bundles over smooth manifolds, e.g. in~\cite{Bouwknegt:2003vb,Evslin:2008zm}.

Next, we consider these backgrounds in the context of type II string theory compactification and find that compactification on $X_5$ is T-dual to compactification on $\Ttor^4/G \times S^1$ equipped with a topologically non-trivial flat $B$-field gerbe.  This duality belongs to a class of T-dual theories, where the geometry is a (possibly singular) principal circle bundle $P_g \to M$, while the choice of gerbe is encoded in a second circle bundle $P_b \to M$.  The product bundle $P_g \times_{M} P_b \to M$ is then the correspondence space~\cite{Bouwknegt:2003vb} for T-duality along the circle:  the action of T-duality exchanges the two circle bundles, so that the geometric fibration data is exchanged with the flat gerbe data.  

  At the level of the orbifold CFT this illustrates, along the lines described in~\cite{Sharpe:2003cs}, that the gerbe data gives a natural geometric meaning to certain shift orbifold phases.  One example of such phases is a choice of discrete torsion~\cite{Vafa:1986wx,Sharpe:2000ki}, but these are absent in all of our examples.  As emphasized in~\cite{Sharpe:2003cs}, the gerbe encodes more than the choice of discrete torsion, and our work explores the physical consequences of this additional data.  
  
The plan of the rest of the paper is as follows.  We begin with a discussion of the geometry of $X_5$ and construct the massless spectrum for heterotic compactification on $X_5$ from a spacetime point of view.  Next we study of the details of the heterotic worldsheet and the duality between compactification on $X_5$ and that on $\Ttor^4/G \times S^1$.  We then turn to the type II string, where we study principal circle bundles over $\Ttor^4/G$ with flat gerbe structure and explore T-dualities between them.
 We end with a summary and discussion of future directions.  Several technical discussions are given in the appendix.

\section*{Acknowledgements}  IVM and MR would like to thank the Institut Henri Poincar{\'e} and its Research in Paris program for hospitality and support while this work was being completed; IVM's work is also supported in part by NSF grant PHY-1914505, and he also thanks the IPhT for additional support during his visit.  IVM also thanks the CTP at Queen Mary University of London for hospitality during the last stages of this project.  RM is partially supported by ERC grants 772408-Stringlandscape and 787320-QBH Structure.  We thank R.~Field, A.~Font, B.~Fraiman, H. Parras de Freitas, M.R.~Plesser, S.~Sethi, and E.~Sharpe for useful communications and conversations, and we are especially grateful to P.~Vaudrevange, M.~Ratz, and S.~Ramos for sharing the classification results for symmetric orbifolds of $\Ttor^6$  with $N=2$ spacetime supersymmetry that they obtained as part of the work in~\cite{Fischer:2012qj,Fischer:2013qza}.

\section{Compactification on flat manifolds: a spacetime perspective} \label{s:flatspacetime}

\subsection{Essentials of compact flat geometries} \label{ss:X5geom}
A compact flat Riemannian manifold $X$ of dimension $n$ has $\R^{n}$ as its universal cover, so that $X = \R^n/\pi_1$, where $\pi_1$ is the fundamental group of $X$.   It is a classic result (reviewed in, for example, ~\cite{Charlap:1986fcm}) that $\pi_1$ is a discrete and co-compact\footnote{A subgroup $H \subset G$ of a Hausdorff topological group $G$ is said to be co-compact if the coset $G/H$ is compact in the quotient topology.} subgroup of $\R^n \rtimes \GO(n)$, the isometry group of $\R^n$, with the following properties: $\pi_1$ is discrete, contains no finite subgroups, and it contains a free abelian normal subgroup $N \subset  \pi_1$ 
generated by $n$ linearly independent translations, and $G=\pi_1/N$ is a finite group.  $G$ is the Riemannian holonomy group of $X$, and it can be shown that every finite group $G$ can be the holonomy of some compact flat Riemannian manifold~\cite{Auslander:1957hg}.\footnote{It is also possible to show that a finitely generated group $\pi_1$ is the fundamental group of some compact flat manifold if and only if $\pi_1$ has no finite subgroups, and there is a normal subgroup $N \subset \pi_1$ that is maximal abelian such that $\pi_1/N$ is a finite group.}

Since $\pi_1$ is a crystallographic group, it is possible to classify all compact flat manifolds of dimension $n$.   Reviews of the relevant computational tools and algorithms are given in~\cite{Opgenorth:1998cat,Fischer:2012qj}.  As one might expect, the classification becomes fairly elaborate for large $n$, and, as far as we are aware, has only been carried out for $n \le 6$.   Fortunately, for our purposes of finding supersymmetric compactifications, we need only consider a much smaller class of manifolds.

The reason for the restriction is that while the spaces are flat, they in general have non-trivial characteristic classes despite vanishing curvature.  There are classification results, complete through dimension $n\le 6$, of compact flat spin manifolds~\cite{Lutowski_2015}, as well as of compact flat K\"ahler manifolds~\cite{Dekimpe:2009kfm}, which are defined to be manifolds with $\pi_1 \subset \C^{n/2} \rtimes\GU(n/2)$.  The latter work also characterizes the compact flat K\"ahler manifolds with holonomy $G \subset \SU(n/2)$, i.e. the flat Calabi-Yau manifolds.\footnote{In this definition of ``Calabi-Yau'' we allow for both $G \subset \SU(n/2-1)$ and $b_1(X) \neq 0$---properties that are often explicitly excluded by definition in discussion of Calabi-Yau $3$-folds.}  
The results of the classification include the following:
\begin{enumerate}
\item for $n=4$ there are $8$ topological types of K\"ahler manifolds, with $G\in\{1,\Z_2,\Z_3,\Z_4,\Z_6\}$, and the only Calabi-Yau manifold is $\Ttor^4$ with $G=1$;
\item for $n=6$ there are $174$ topological types of flat compact K\"ahler manifolds, and $13$ types of Calabi-Yau manifolds with $G\in \{1,\Z_2,\Z_3,\Z_4,\Z_6, \Z_2\times\Z_2, D_8\}$, where $D_8$ is the dihedral group of order $8$.
\end{enumerate}
The Hodge numbers of these manifolds are also determined in~\cite{Dekimpe:2009kfm}.  For example, the authors obtain the following result for Calabi-Yau $X$ with $b_1(X) = 2$:
\begin{align}
\begin{matrix}
		G 	  && 1 && \Z_2 && \Z_3 && \Z_4 & \Z_6 \\
\# \text{of types} && 1 && 2 	&& 2     && 2 	  & 1
\end{matrix}
\end{align}
Each of these manifolds can be used as a background for supersymmetric compactifications of string theory or supergravity, and although the class is not large, the relative simplicity makes such compactifications interesting, especially, as we will see, in the realm of stringy geometry where the volume of some cycle becomes comparable to the string scale.

From the classification results we see that $n=6$ is the smallest dimension with non-trivial compact flat Calabi-Yau manifolds, and these have a natural sub-class, where $X$ is isometric to $X = X_5\times S^1$, and $X_5$ is compact flat manifold with $G = \Z_N$ for $N \in\{1,2,3,4,6\}$.  It is easy to construct examples of such $X_5$, as we now discuss.

Let $(z_1,z_2;\theta)$ denote coordinates on $\Ttor^2\times \Ttor^2 \times S^1$ with the familiar identifications
\begin{align}
(z_1,z_2;\theta) \sim (z_1 + m_1 + n_1 \tau_1, z_2 + m_2 + n_2\tau_2, \theta+2\pi m_3) 
\end{align}
for $m_{1,2,3}, n_{1,2} \in \Z$, and with $\tau_1,\tau_2$ being the complex structure parameters for the $\Ttor^2$ factors.  Every $\Ttor^2$ has a reflection symmetry $\Z_2 : z\mapsto -z$; when $\tau = i$ the $\Ttor^2$ admits a $\Z_4$ action $\Z_4:  z\mapsto i z$, and when $\tau = e^{2\pi i/6}$ there is a $\Z_{3}$ action $z\mapsto e^{2\pi i/3} z$.  In this latter case there is also a $\Z_6 = \Z_2 \times \Z_3$ action $z\mapsto e^{2\pi i /6}z$.  Using these symmetries (after tuning the $\tau_i$ appropriately), we obtain some familiar $\Z_N$ quotients of $\Ttor^4$ that are consistent with its hyper-K\"ahler structure:
\begin{align}
\label{eq:HyperKaehlerOrbifolds}
(z_1,z_2) \mapsto (\zeta_N z_1, \zeta_{N}^{-1} z_2)~,
\end{align}
where $\zeta_N = e^{2\pi i/N}$.  Of course such a quotient leads to a singular orbifold $\Ttor^4/\Z_N$, and the corresponding CFTs have been studied extensively in the literature, e.g.~\cite{Aldazabal:1997wi,Stieberger:1998yi,Honecker:2006qz}, with a focus on $6$ dimensional heterotic vacua and aspects of string duality.  All of these can be understood as degenerations of a K3 surface~\cite{Aspinwall:1995zi,Aspinwall:1996mn}, giving a connection between these orbifold CFTs and compactification on a smooth large radius geometry.

Now it is a simple matter to construct our smooth $X_5$.  For each $N \in\{2,3,4,6
\}$ (and tuning $\tau_i$ appropriately for $N>2$), we take the $G = \Z_N$ action on $X_5$ with generator
\begin{align}
g (z_1,z_2,\theta) = \left(\zeta_N z_1, \zeta_N^{-1} z_2, \theta + \ff{2\pi}{N} \right)~.
\end{align}

The Calabi-Yau flat geometries naturally fit into a larger class of spaces, the supersymmetric orbifolds of $\Ttor^6$ classified in~\cite{Fischer:2012qj}.  The authors of~\cite{Fischer:2012qj} provided us with the list of orbifolds that preserve $N=2$ spacetime supersymmetry, and when we further restrict attention to freely-acting orbifolds that leave one $S^1$ invariant, we recover exactly the class of geometries just described, and these geometries will play a key role in our investigation.

\subsubsection*{Two fibration structures}
There are two useful perspectives on the geometry of $X_5$ as a fibration.  First, we can think of the projection $\mu_{\text{cir}} :  X_5 \to S^1$, with $\mu_{\text{cir}}(z_1,z_2,\theta) = N\theta$,  where the fiber $\mu_{\text{cir}}^{-1} (\varphi) \simeq \Ttor^4$, and the fibers $\mu_{\text{cir}}^{-1}(\varphi)$ and $\mu_{\text{cir}}^{-1}(\varphi+2\pi)$ are related by the $\Z_N$ action on $\Ttor^4$.  This structure will be useful when we think of compactification on $X_5$ in two steps:  first, we compactify string theory on $\Ttor^4$, and then we compactify the resulting theory further on an additional circle with a non-trivial holonomy for the $G = \Z_N$ action.

Second, we have the projection $\mu_{\text{orb}} :  X_5 \to \Ttor^4/\Z_N$, where the map simply forgets the last coordinate and maps the torus coordinates to the corresponding equivalence class:
$\mu_{\text{orb}} (z_1,z_2,\theta) = [z_1,z_2]$.  In this case, the fiber is a circle, but with an identification that jumps at the fixed points of the $\Ttor^4/\Z_N$ orbifold:  for a generic point in the base we identify $\theta \sim \theta+2\pi$, while at each fixed point $p\in \Ttor^4/\Z_N$ we have $\theta\sim\theta +2\pi/n(p)$, where $n(p)$ is the order of the stabilizer subgroup of the point $p$.

\subsection{Heterotic compactification to six dimensions: set up}
We begin our study of heterotic compactification on $X_5$ with a review of the well-known construction of heterotic theory compactified on $\Ttor^4$.  This compactification yields a $d=6$ (1,1) supergravity theory with massless content consisting of the (1,1) supergravity multiplet and a number of vector multiplets.  Each vector multiplet contains $4$ scalars that transform in the adjoint of the gauge algebra.  At a generic point in the scalar moduli space $\Gr(20,4)/\GO(\Gamma_{20,4})$ the gauge algebra is $\Lu(1)^{\oplus 20}$.\footnote{Here $\Gr(20,4)$ is the coset $\SO(20,4,\R)/\SO(20,\R)\times \SO(4,\R)$, and $\GO(\Gamma_{20,4})$ is the group of lattice isomorphisms of the even self-dual lattice $\Gamma_{20,4}$, often written as $\GO(20,4,\Z)$.}  At special points the gauge algebra can be enhanced, and the enhanced symmetries and the corresponding loci in the moduli space have been recently studied in~\cite{Font:2020rsk}.    For our purposes it will be sufficient to consider the locus where the gauge algebra is $\Le_8 \oplus \Lu(1)^{\oplus 4} \oplus \Le_8$:  this locus has a standard RNS worldsheet realization, which will make it easy to describe the constructions we wish to consider.  In this section we will summarize some of the relevant details, and in the next section we will apply them to some classic examples.

We work in light-cone gauge on a Euclidean worldsheet with coordinates $z,\zb$, with worldsheet supersymmetry on the right (anti-holomorphic) side of the string.  In addition to the degrees of freedom for the Minkowski directions, the internal CFT consists of the scalar fields $\Phi^i(z,\zb)$ and their right-moving Majorana-Weyl superpartners, $8$ left-moving Weyl fermions, and a level $1$ $\Le_8$ current algebra for the ``hidden $\GE_8$'' that will play a spectator role in our analysis.

We find it convenient to break up the left-moving fermions into $2$ Weyl fermions $\gamma^{1,2}(z)$ and their conjugates $\gammab^{1,2}(z)$, and $6$ more Weyl fermions $\xi^I(z)$, $\xib^I(z)$; similarly, we organize the right-moving fermions into a Weyl pair  $\psi^{1,2}(\zb)$, with conjugates $\psib^{1,2}(\zb)$.  This gives a decomposition of the corresponding level $1$ current algebras
\begin{align}
\so(16)_L &\supset \so(12)_L \oplus \su(2)_{L} \oplus \su(2)_{L}'~, &
\so(4)_R &\simeq \su(2)_{R} \oplus \su(2)_{R}'~.
\end{align}
The fermions transform in the following representations.  On the holomorphic side
\begin{align}
\xi &\in (\rep{12},\rep{1},\rep{1})~,&
\gamma,\gammab & \in (\rep{1},\rep{2},\rep{2})~,
\end{align}
while on the anti-holomorphic side we have $\psi,\psib\in (\rep{2},\rep{2})$ of $\su(2)_R\oplus\su(2)'_R$.

This structure is a special case of a (0,4) SCFT necessary~\cite{Banks:1988yz} for the preservation of (1,0) supersymmetry in $\R^{1,5}$.  As our ultimate aim will be study compactifications with $8$ supercharges, we will now describe some features in this more general setting, focusing on the identification of states in the worldsheet theory with massless fermions in spacetime.

The connection is established through the current algebra.  The key ingredient is $\su(2)_R$, which we identify as the R-symmetry of the N=4 superconformal algebra (SCA), and it contains $\Lu(1)_R$, an R-symmetry for an N=2 subalgebra of the N=4 SCA with current $\Jb$ and operator charges labeled by $\qb$.  We will also choose $\Lu(1)_L \subset \su(2)_L$ with current $J$ and charges $q$ to label our states, and we similarly define the currents and charges $J',q'$ for $\Lu(1)'_L \subset \su(2)'_L$ and $\Jb',\qb'$ for $\Lu(1)'_R \subset \su(2)'_R$.   The conserved charges $J_0$ and $\Jb_0$  associated to $\Lu(1)_L \oplus\Lu(1)_R$ give simple expressions for the internal left- and right-fermion numbers that are necessary for the GSO projections: we have $(-1)^{F_{\gamma}} = e^{i \pi J_0}$, while $(-1)^{F_\psi} = e^{i\pi\Jb_0}$.

We can now use the familiar rules---see e.g. \cite{Polchinski:1998rr,Melnikov:2019tpl}---to identify worldsheet states with massless multiplets in spacetime.  
\begin{enumerate}
\item The identity operator of the internal CFT gives rise to a (1,0) supergravity multiplet and a (1,0) tensor multiplet.\footnote{Details of the multiplet structure can be found in the recent pedagogical review~\cite{Lauria:2020rhc}.}
\item The spacetime gauge bosons arise in two ways:  every GSO-even holomorphic current gives rise to a spacetime gauge boson, and every anti-holomorphic operator with weight $\hb = 1/2$  (i.e. a free fermion) gives rise to an abelian gauge boson.  The latter, when present, complete the (1,0) gravity multiplet and the (1,0) tensor multiplet to a (1,1) supergravity multiplet, while each of the former corresponds to a (1,0) vector multiplet.

In all of our examples the holomorphic current algebra will be of the form $\Le_8\oplus \Lu(1)^{\oplus k} \oplus \Lg$, and in most of our examples $\Lg \supset \Le_7\oplus\Lu(1)_L'$.  Note that the ``linearly realized'' currents of $\so(12) \oplus \su(2)_L$ are completed to $\Le_7$ by additional currents from the left-moving Ramond sector in the $(\rep{32},\rep{2})$ representation.

In $\Ttor^4$ compactification the $\Lu(1)'_L$ current is enhanced to $\su(2)_L'$, and there are $h=1/2$ holomorphic operators transforming in $\rep{2}$ of $\su(2)_L'$, and additional currents
\begin{align}
(\rep{32},\rep{2},\rep{1}) \oplus(\rep{12},\rep{2},\rep{1}) \oplus
(\rep{32'},\rep{1},\rep{2}) \oplus (\rep{12},\rep{1},\rep{2})~,
\end{align}
that complete $\Lg$ to $\Lg = \Le_8$.

Finally, the $\Lu(1)^{\oplus k}$ factor arises from additional currents neutral with respect to $\Lg$.  For example in $\Ttor^4$ compactification, $k=4$ with the currents $i\p \Phi^i$.

\item The remaining spacetime massless fields reside in (1,0) $\ff{1}{2}$--hypermultiplets.  These are in one to one correspondence with the $\overline{\text{NS}}$ chiral primary states with $\qb = 1$ and holomorphic weight $h=1$.  Their gauge transformations are determined by the left-moving sector: each worldsheet state with $q=0$ leads to an $\Le_7$--neutral $\ff{1}{2}$--hypermultiplet, while each worldsheet state with $q=1$ leads to a state transforming in $\rep{56}$ of $\Le_7$.

\end{enumerate}
Let us apply these rules to the $\Ttor^4$ example, focusing on the (1,0) vector and hypermultiplets.
The R-charge assignments to the right-moving fermions are
\begin{align}
\begin{matrix}
~	&& \psi^{1} 	&& \psi^{2} 	&& \psib^1 	&& \psib^2 \\
\qb	&&+1		&& +1			&& -1	      	&& -1  \\
\qb'	&&+1		&&-1			&&-1	      		&&+1 
\end{matrix}
\end{align}
We can easily see that the spectrum is consistent with (1,1) spacetime supersymmetry: for every holomorphic current $J$ we have the operators $J \psi^1$ and $J \psi^2$ that together give rise to a (1,0) hypermultiplet transforming in the same way as the current $J$ under the gauge symmetry.

\subsection{$\Ttor^4/\Z_N$ compactifications}
We now turn our attention to (1,0) theories that are built by considering symmetric orbifolds $\Ttor^4/\Z_N$.  In each case, we take the $\Z_N$ action to be just the action described in~(\ref{eq:HyperKaehlerOrbifolds}), extending it to act on the worldsheet fermions in the standard left-right--symmetric fashion:  the generator $g$ of the $\Z_N$ action is set to be
\begin{align}
\label{eq:RNSZNaction}
g = \underbrace{\exp\left[ \ff{2\pi i}{N} J'_0\right]}_{=g_\gamma} ~~\underbrace{ \exp\left[ \ff{2\pi i}{N}\Jb'_0 ) \right]}_{=g_\psi}~.
\end{align}
The $\Z_N$ action is diagonally embedded in $\GU(1)_{L}' \times\GU(1)_{R}'$.  For $N=2$ the action is contained in the center of each corresponding $\SU(2)$ factor, so that the $\su(2)_L' \oplus\su(2)_R'$ algebra is left invariant, while for $N>2$ the invariant subalgebra is $\Lu(1)_L' \oplus \Lu(1)_R'$.  We are of course by no means the first to consider these orbifolds---see, for example~\cite{Walton:1987bu,Aldazabal:1997wi,Stieberger:1998yi,Kaplunovsky:1999ia,Honecker:2006qz} and references therein for earlier work and some asymmetric generalizations.  However, we will recall some of the details and give a view that will be useful for what follows.
\subsubsection*{Untwisted massless states}
To describe the untwisted massless states, we just need to apply the projection onto $\Z_N$--invariant states.  We first observe that for all $N>0$ the special anti-holomorphic states with $\hb = 1/2$ are projected out, and therefore the spacetime supersymmetry is reduced.  However, since the projection keeps the $\su(2)_R$ current algebra and leaves the vacuum of the internal theory invariant, we are guaranteed (1,0) supersymmetry.  This makes the preceding discussion of the multiplet structure well-adapted to understand the projection. 

The invariant content of the (1,1) supergravity multiplet is exactly the (1,0) supergravity and tensor multiplets.  To analyze the invariant content of the (1,1) vector multiplets, we use the decomposition
\begin{align}
\text{ (1,1) vector} = \text{(1,0) vector} \oplus \text{ (1,0) hyper}~.
\end{align}
The $\Z_N$ action on each multiplet is a combination of an $\SU(2)'_L$ gauge symmetry, an $\SU(2)'_R$ R-symmetry, as well as $G_{\text{ab}} \simeq \Z_N$, which acts on the abelian (1,1) vector multiplets corresponding to the $\Lu(1)^{\oplus 4}$ factor according to~(\ref{eq:HyperKaehlerOrbifolds}).  To describe that action, we package the $4$ vector multiplets into two complex combinations that transform according to
\begin{align}
\rep{1}_{\zeta_N} \oplus \rep{1}_{\zeta_N^{-1}}~
\end{align}
under the action of $G_{\text{ab}}$.


Consider first the (1,1) vector multiplets corresponding to the hidden $\Le_8$ symmetry. While the (1,0) vector submultiplets of these are left invariant, every $\ff{1}{2}$-hyper is projected out, so we are left with just the $(1,0)$ vector multiples of $\Le_8$.

Next, we examine the $(1,1)$ abelian vectors: now each $(1,0)$ vector is projected out, but because of the additional action in $\SU(2)_R'$, the $(1,0)$ hypermultiplet content is more interesting:  under $\SU(2)_{R}'\times G_{\text{ab}}$ the $\ff{1}{2}$--hypermultiplets transform as
\begin{align}
\rep{2}_{\zeta_N} \oplus \rep{2}_{\zeta_N^{-1}}~,
\end{align}
or decomposing further with respect to $\GU(1)_{R}'$, as
\begin{align}
\rep{1}_{1,\zeta_N} \oplus \rep{1}_{-1,\zeta_N} \oplus \rep{1}_{1,\zeta_N^{-1}} \oplus \rep{1}_{-1,\zeta_N^{-1}}~.
\end{align}
Thus, if $N=2$, then all of these are invariant, while if $N>2$ half of these are invariant.  So, we find $4$ neutral hypers for $N=2$ and $2$ neutral hypers for $N>2$.

To discuss the $\Le_8$ vectors, we decompose
\begin{align}
\Le_8 &\supset \Le_7 \oplus \Lu(1)'_L \nonumber\\
\rep{248} & = \rep{133}_0 \oplus \rep{1}_0 \oplus \rep{56}_{+1} \oplus \rep{56}_{-1} \oplus\rep{1}_{+2} \oplus \rep{1}_{-2}~.
\end{align}
From this we see that for $N>2$ the invariant $(1,0)$ vectors transform in the adjoint of $\Le_7\oplus\Lu(1)'$, while for $N=2$ we find the adjoint of $\Le_7\oplus\su(2)'_L$.  To obtain the invariant $\ff{1}{2}$-hypers, we tensor this with $\rep{2}$ of $\su(2)_{R}'$, or equivalently with $\rep{1}_{+1} \oplus\rep{1}_{-1}$ of $\Lu(1)_{R}'$:
\begin{align}
\rep{248} & = \rep{133}_{0,+1} \oplus \rep{133}_{0,-1} \oplus \rep{1}_{0,+1} \oplus\rep{1}_{0,-1} \nonumber\\
&\quad \oplus \rep{56}_{+1,+1} \oplus \rep{56}_{+1,-1} \oplus \rep{56}_{-1,+1} \oplus \rep{56_{-1,+1} }
\nonumber\\
&\quad \oplus\rep{1}_{+2,+1} \oplus\rep{1}_{+2,-1}\oplus \rep{1}_{-2,+1}\oplus \rep{1}_{-2,-1}~.
\end{align}
The invariant $\ff{1}{2}$-hypers are those with $q'+\qb' = 0\mod N$. The result is the following invariant spectrum from the untwisted sector (as promised, we ignore the hidden $\Le_8$):
\begin{align}
\label{eq:TorusOrbifoldUntwisted}
\Z_N && \text{gauge symmetry} && \text{hypers} \nonumber\\
\Z_2 &&\Le_7 \oplus \su(2)_L'&& \rep{1}_0^{\oplus 4} \oplus (\rep{56},\rep{2}) \nonumber\\
\Z_3 && \Le_7 \oplus \Lu(1)_L'  && \rep{1}^{\oplus 2}_0 \oplus \rep{1}_2 \oplus \rep{56}_1 \nonumber\\
\Z_{4,6} &&\Le_7 \oplus \Lu(1)_L'  && \rep{1}^{\oplus 2}_0 \oplus \rep{56}_1
\end{align}

\subsubsection*{Twisted sector contributions}
A simple approach to work out the twisted sector contributions is to model each fixed point of the orbifold by $\C^2/\Z_N$ and use free field techniques to calculate the quantum numbers of states in the twisted sector.  It is then not too difficult to read off the twisted sector states that yield spacetime massless states.  We will not have need for details of the construction, so we will simply quote the results from the literature~\cite{Aldazabal:1997wi,Honecker:2006qz}.  The twisted sectors make no contribution to the gauge symmetry, but they do produce additional hypermultiplets arranged as follows:
\begin{align}
\begin{array}{ccl}
\label{eq:TorusOrbifoldTwisted}
\Z_N &\hspace{1cm}&  \text{twisted hypers} \\[2mm]
\Z_2 &\hspace{1cm}&(\rep{2},\rep{1})^{\oplus 32} \oplus(\rep{1},\rep{56})^{\oplus 8}\\[2mm]
\Z_3 &\hspace{1cm}&  \rep{56}^{\oplus 9}_{-1/3}\oplus \rep{1}^{\oplus 45}_{2/3} \oplus \rep{1}^{\oplus 18}_{-4/3}\nonumber\\[2mm]
\Z_{4} &\hspace{1cm}&  \rep{56}_{-1/2}^{\oplus 4} \oplus \rep{1}_{3/2}^{\oplus 8} \oplus \rep{1}_{1/2}^{\oplus 24} \oplus \rep{56}_0^{\oplus 5} \oplus \rep{1}_1^{\oplus 32}\\[2mm]
\Z_{6} &\hspace{1cm}&  
 \rep{1}^{\oplus 8}_{1/3} \oplus \rep{1}^{\oplus 2}_{-5/3}  \oplus \rep{56}_{-2/3} \oplus\rep{1}^{\oplus 22}_{2/3} \oplus\rep{1}^{\oplus 10}_{-4/3} \oplus \rep{56}^{\oplus 5}_{-1/3} \oplus \rep{1}_1^{\oplus 22} \oplus \rep{56}_0^{\oplus 3}
\end{array}
\end{align}
The reader can check that in all cases the six-dimensional anomaly cancelation condition $N_{\text{hyper}} - N_{\text{vector}} = 244$ is satisfied.

Each of these theories has a Higgs branch: we can Higgs the $\su(2)_{L}'$ for $N=2$ and the $\Lu(1)_L'$ for $N>2$ to obtain a spectrum with gauge algebra $\Le_7$ and hypermultiplets in $(\rep{56})^{\oplus 10}$---which is exactly the massless spectrum of heterotic compactification on K3 with standard embedding.  We expect this result since each of these singularities is a degeneration limit of the K3 surface.\footnote{The limit in the non-linear sigma model's moduli space is subtle due to a choice of $B$-field on the collapsing two-cycles of the degenerating K3 surface: see~\cite{Aspinwall:1995zi,Aspinwall:1996mn} for further discussion.}  

\subsection{Compactification on $\Ttor^4/\Z_N\times S^1$}
Having obtained the $d=6$ theory, it is a simple matter to compactify further on a circle.  If we treat the circle as completely decoupled, then the massless spectrum is obtained by standard Kaluza-Klein reduction~\cite{Lauria:2020rhc}:
\begin{enumerate}
\item the reduction of the (1,0) supergravity and tensor multiplets leads to a (minimal) $d=5$ supergravity multiplet and $2$ abelian vector multiplets\footnote{One of these is obtained by dualizing a $d=5$ abelian tensor multiplet.};
\item each $d=6$ vector multiplet in representation $\rep{r}$ of the gauge algebra reduces to a $d=5$ vector multiplet, which now has a real scalar transforming in the same representation;
\item similarly, each $d=6$ hypermultiplet reduces to a $d=5$ hypermultiplet in the same representation of the gauge algebra.
\end{enumerate}
All in all, the gauge algebra is now $\Lu(1)^{\oplus 2} \oplus \Le_8 \oplus \Lg$, where the first factor is due to the vector multiplets obtained from the reduction of supergravity and tensor (1,0) multiplets, and the hypermultiplet spectrum is unmodified.

The new feature in $d=5$ is the existence of a Coulomb branch.  Giving a generic set of expectation values to the scalars in the vector multiplets breaks the gauge group to $\Lu(1)^{\oplus  18}$ and leaves $4$ neutral hypermultiplets in the $\Z_2$ orbifold and just $2$ neutral hypermultiplets for $\Z_{3,4,6}$.  Observing that under $\Le_7 \supset \Le_6\oplus\Lu(1)$
\begin{align}
\rep{56} = \rep{27}_{+1} \oplus\brep{27}_{-1} \oplus\rep{1}_{+3}\oplus \rep{1}_{-3}~,
\end{align}
we see that for the $\Z_{3,4,6}$ compactifications there is a Coulomb branch with unbroken $\Lg = \Lu(1)^{2} \oplus \Le_8 \oplus\Le_6 \oplus\Lu(1)'$ where all of the twisted sector matter is lifted, and there are $2$ neural massless hypermultiplets.  A special feature of the $\Z_3$ example is that it is possible to lift all of the twisted sector states by just going on the $\Lu(1)'$ Coulomb branch, where the full gauge algebra $\Lg$ is preserved.

\subsection{Compactification on $X_5$} \label{ss:X5compactification}
Having reviewed the well-known compactifications on $\Ttor^4/\Z_N\times S^1$, we now consider compactification on $X_5$.  From the worldsheet point of view, the new ingredient relative to the preceding discussion is a $\Z_N$ shift orbifold of the circle.  By itself this theory is easy to understand:  starting with a CFT for a circle of radius $r$, the shift orbifold is a CFT of radius $r/N$~\cite{Ginsparg:1988ui}.  In the full quotient CFT for $X_5 = (\Ttor^4\times S^1)/\Z_N$, we just need to combine the twisted sectors of the $\Ttor^4/\Z_N$ and $S^1/\Z_N$ CFTs and impose the orbifold projection.  With a view to later developments, we will first describe the construction of the $S^1$ shift orbifold in some detail.

\subsubsection*{The $\Z_N$ shift orbifold of the circle} \label{ss:shiftorbifold}
Consider a compact $c=1$ boson at a generic radius $r$.  Splitting the worldsheet field $\Phi(z,\zb)$ into left- and right-moving components, we have the defining OPEs
\begin{align}
\Phi_L(z) \Phi_L(w) &\sim - \ff{1}{r^2} \log(z-w)~,&
\Phi_R(\zb) \Phi_R(\wb) & \sim -\ff{1}{r^2} \log(\zb-\wb)~.
\end{align}
The theory enjoys a Kac-Moody $\GU(1)^{\text{shift}}_L\times\GU(1)^{\text{shift}}_R$ symmetry with currents
\begin{align}
J &= i r \p \Phi_L~, &
\Jb &  = i r \pb \Phi_R~,
\end{align} 
and the (Kac-Moody) primary states $|p\ra$ are labeled by the momentum and winding modes $n,w\in \Z$.  More precisely, 
\begin{align}
p = w\be + n \be^\ast \in \Gamma_{1,1}~,
\end{align} 
with $\Gamma_{1,1} \subset\R^{1,1}$ the even self-dual lattice spanned by two lattice vectors $\be$ and $\be^\ast$ satisfying $\be.\be = \be^\ast.\be^\ast=0$ and $\be.\be^\ast = 1$.  Here $a.b$ denotes the inner product on $\R^{1,1}$ induced by the bilinear pairing on the lattice.  Note that for our discussions of lattices here and in what follows, we will take $\R^{n_L,n_R}$ to have Lorentzian metric 
\begin{align}
\eta & = \begin{pmatrix} -\iden_{n_L}  & 0 \\ 0 & +\iden_{n_R} \end{pmatrix}~.
\end{align}
Our theory has a one-dimensional moduli space, which we can think of as a specification of a spacelike $1$-plane $\Pi \subset \R^{1,1}$.  We can always choose a basis vector for $\Pi$
\begin{align}
\bpi & = \be + r^2 \be^\ast~,
\end{align}
so that $\bpit = \be - r^2 \be^\ast$ spans the orthogonal complement $\Pi_\perp$. 

With that preparation, we write the operator corresponding to the state $|p\ra$ as
\begin{align}
\cV_p & = ~:~ \bC(p) \exp\left[ \ff{i}{\sqrt{2}} (\bpi .p) \Phi_R(\zb) + \ff{i}{\sqrt{2}} (\bpit.p) \Phi_L(z) \right]~:~.
\end{align}
$\bC(p)$ is an operator constructed from the momentum zero modes that we will discuss further below; this ``cocycle operator'' is necessary to ensure proper commutation relations of the vertex operators.   Using the OPEs, it is now straightforward to see that $\cV_p$ carries $\GU(1)^{\text{shift}}_L \times\GU(1)^{\text{shift}}_R$ charges
\begin{align}
q_L^{\text{sh}} & = \frac{\bpit.p}{r\sqrt{2}}~,&
q_R^{\text{sh}} & = \frac{ \bpi.p}{r\sqrt{2}}~
\end{align}
and weights 
\begin{align}
\label{eq:circleweights}
h_L & = \ff{1}{2} (q_L^{\text{sh}})^2= \ff{1}{4r^2} \left(n-wr^2\right)^2~,&
h_R &  = \ff{1}{2} (q_R^{\text{sh}})^2 = \ff{1}{4r^2} \left(n+wr^2\right)^2~.
\end{align}
The spin of the operator is
\begin{align}
s(p) = h_L - h_R = -\ff{1}{2} p.p = -nw~.
\end{align}
Note that in these conventions the self-dual radius is $r=1$, with the T-duality map being $r\mapsto 1/r$ and $(n,w) \mapsto (w,n)$.

With this preparation, we define the action of the shift symmetry $\Z_N^{\text{shift}}$, taking its generator $g_{\text{sh}}$ to act as
\begin{align}
g_{\text{sh}} |p\ra & = e^{2\pi i n /N} |p\ra~.
\end{align}
Since the momentum quantum number $n$ is conserved, this is clearly a symmetry of the spectrum, of the OPE, and of the correlation functions.  Moreover, we can represent this action in terms of the conserved shift currents (or, rather, the corresponding conserved charges):
\begin{align}
g_{\text{sh}} & = e^{ \frac{2\pi i}{N} Q}~,
\end{align}
with the charge $Q$ given by
\begin{align}
Q =  \ff{r}{\sqrt{2}} \left(J_{L,0}^{\text{sh}} + J_{R,0}^{\text{sh}} \right)~.
\end{align}
The orbifold projection is then onto states with $Q \in N \Z$.

This is a key simplification in the orbifold analysis, since it allows us to directly construct the twisted Hilbert space---see~\cite{Melnikov:2019tpl} for a recent pedagogical discussion.\footnote{A recent discussion of the more general situation is given in~\cite{Robbins:2019zdb}.}  To carry this out, we find a twist field of the form
\begin{align}
\Sigma_k(z,\zb) = \exp\left[-i \ff{r^2}{\sqrt{2}}\tau_k \Phi_R -i\ff{r^2}{\sqrt{2}} \taut_k \Phi_L\right]~,
\end{align}
where the parameters $\tau_k$ and $\taut_k$ are chosen so that the OPE of $\cV_p(z,\zb)$ and $\Sigma_k(z,\zb)$ has the correct monodromy, i.e. so that under a continuous rotation $z \to e^{i\theta} z$, we obtain
\begin{align}
\cV_p(e^{2\pi i} z,e^{-2\pi i}\zb) \Sigma_k(0) = e^{\ff{2\pi i kn}{N}} \cV_p(z,\zb)~. \Sigma_k(0)
\end{align}
The field $\Sigma$ then gives the $k$-th twisted ground state $|0;k\ra = \lim_{z,\zb\to 0} \Sigma(z,\zb) |0\ra$, and the full Hilbert space in the $k$-th twisted sector is constructed from $|p;k\ra$---obtained by acting further with the $\cV_p$ on the $|0;k\ra$---by acting with all possible oscillators.  

In our example a quick computation shows that choosing
\begin{align}
\tau_k &= -\ff{k}{N}~, &
\taut_k & = \ff{k}{N}~
\end{align}
produces the correct monodromy.  Moreover, it is straightforward to read off the shifts in the weights and charges due to the twist, i.e. the weights and charges of the state $|p;k\ra$:
\begin{align}
q^{\text{sh}}_L(p;k) & = \frac{\pit.p-\taut_k r^2}{r\sqrt{2}}~,&
q^{\text{sh}}_R(p;k) & = \frac{\pi.p -\tau_k r^2}{r\sqrt{2}}~,
\end{align}
and the weights are
\begin{align}
h_L(p;k) & = \ff{1}{2} (q_L^{\text{sh}}(p;k))^2~,&
h_R(p;k) &  = \ff{1}{2} (q_R^{\text{sh}}(p;k))^2~.
\end{align}
Not only do we know the twisted Hilbert spaces, but we also know how to implement the projection onto invariant states:  we just need to project onto states with
\begin{align}
Q(p;k) = \frac{r}{\sqrt{2}} \left( q_L^{\text{sh}}(p;k) + q_R^{\text{sh}}(p;k) \right) \in N \Z~.
\end{align}
For our circle shift orbifold it is easy to check that $Q(p;k) = Q(p)$, and also that the spin satisfies
\begin{align}
s(p;k) = h_L(p;k) - h_R(p;k) = s(p) -\frac{k}{N} Q(p;k)~, 
\end{align}
so that $\Z_N$--invariant states are guaranteed to have integer spin.

Carrying out the construction for the $N-1$ twisted sectors, we find the expected structure:  the projection sets the momentum modes to be valued in $N\Z$, as is consistent with a circle of radius $r/N$, while the $k$-th twisted sector adds in the winding modes with $w \in \ff{k}{N} + \Z$, which are the ``extra'' winding modes for a circle of radius $r/N$ relative to that of radius $r$.  

\subsubsection*{The $X_5$ orbifold}
Having understood the shift orbifold in detail, we could easily construct the full partition function for the $X_5$ CFT.  The only modification to our previous discussion of the $\Ttor^4/\Z_N$ compactification is to treat the right-moving superpartner of $\Phi(z,\zb)$ as part of the internal CFT rather than belonging to the CFT describing the $\R^{1,5}$ degrees of freedom.\footnote{There are some subtleties in applying the usual RNS rules in odd-dimensional compactification, but they can be easily avoided by introducing an additional spectator circle with its superpartner; see e.g.~\cite{Melnikov:2017yvz}.}

If our interest is in the massless spectrum, then the effect of the extra shift at generic radius is simple to understand: the untwisted massless states are exactly those of the $\Ttor^4/\Z_N\times S^1$ CFT, while all of the twisted sector states are massive.  This observation was the starting point for our study, since it suggests a close relationship between the two theories.

\subsubsection*{A spacetime picture}
There is another way to think about the $X_5$ compactification, inspired by the $\mu_{\text{cir}}$ fibration described in section~\ref{ss:X5geom}
\begin{equation*}
\begin{tikzcd}
\Ttor^4 \arrow[r,hook] & X_5 \arrow[d, "\mu_{\text{cir}}"]  \\ & S^1
\end{tikzcd}~.
\end{equation*}
We start in the $d=6$ (1,1) obtained by heterotic compactification on $\Ttor^4$. The worldsheet symmetries we identified in our discussion of the $\Ttor^4/\Z_N$ orbifold are interpreted as spacetime gauge symmetries, and in particular there is a discrete gauge symmetry $G = \Z_N$ which is a subgroup
\begin{align}
G \subset G_{\text{ab}} \times \SU(2)'_L \times \SU(2)'_{R}~,
\end{align}
the latter action being part of the spacetime R-symmetry group of the $d=6$ (1,1) theory.  Given the presence of this discrete spacetime gauge symmetry, when we compactify the theory on a circle with coordinate $\theta\sim\theta+2\pi$, we can choose to turn on a holonomy for $G$.  That is, we modify the periodicity conditions on the fields as
\begin{align}
\phi(x,\theta+2\pi) = g \cdot \phi(x,\theta)~.
\end{align}
The holonomy will have the effect that only $G$--invariant fields will have zero modes in the $\theta$ expansion, so that the low energy theory obtained in this way will have the same massless spectrum as compactification on $X_5$.  Because $G$ involves an $\SU(2)'_R$ action, half of the gravitinos will be lifted in the process, reducing supersymmetry from $16$ to $8$ supercharges.  

Our point of view is that the orbifold construction of the heterotic string on $X_5$ is a UV completion of circle compactification of the spacetime theory with a $G$-holonomy.   Note we are not suggesting a spacetime interpretation relating compactification based on a CFT to that based on the orbifold CFT.  Instead, the relationship is between a compactification based on a freely acting orbifold and a circle compactification of a theory from one dimension higher.  In the appendix we present a toy bosonic string model illustrating this structure for the $\Z_2$ quotient. 

It is clear that the existence of a UV completion is a non-trivial condition because not every global discrete symmetry of the heterotic CFT leads to a modular-invariant orbifold, yet we expect such symmetries to still correspond to gauge symmetries in the spacetime theory.  Thus, there should in general be obstructions to turning on non-trivial holonomy for a discrete spacetime gauge symmetry $G$.  We hope to return to a study of these in the future.

\section{The heterotic worldsheet and duality}
In the previous section we presented two classes of $d=5$ heterotic compactifications that preserve $8$ supercharges.  While we saw that their spectra are closely related---in fact identical when restricted to the massless untwisted states---in general the theories appear to be distinct.  As a stark difference we noted that $\Ttor^4/\Z_N \times S^1$ compactifications can be deformed to $\text{K3}\times S^1$ by going on the Higgs branch, while no such Higgs branch appears to be present in the $X_5$ compactifications.

Nevertheless, we will now show that by enlarging the moduli space to include Wilson lines we can establish an isomorphism between these classes of theories.  Thus, not only are these compactifications connected in moduli space, they are in fact identical.  

\subsection{Bosonic construction} \label{ss:Narain}
Turning on Wilson line parameters is best described in the bosonized construction of the heterotic worldsheet CFT on $\Ttor^d$~\cite{Narain:1985jj,Narain:1986am}.  Grouping the left-moving fermions into $16$ Weyl fermions $\xi^a$, $\xib^a$, so that the currents $J^a =~ : \xi^a \xib^a:$ generate a Cartan algebra of $\Le_8\oplus\Le_8$, we bosonize these as $J^a = i \p \cX^a_L$.
The bosonic OPEs now depend on the metric $g_{ij}$ on $\Ttor^d$:
\begin{align}
\Phi^i_L(z) \Phi^j_L(w) &\sim - g^{ij} \log(z-w)~,&
\Phi^i_R(\zb) \Phi^j_R(\wb) & \sim -g^{ij} \log(\zb-\wb)~, \nonumber\\
\cX_L^a(z) \cX_L^b(w) & \sim - \delta^{ab} \log(z-w)~.
\end{align}
The vertex operators are labeled by $p \in \Gamma_{d+16,d} \subset \R^{d+16,d}$.  The lattice isomorphism $\Gamma_{d+16,d} \simeq (\Gamma_{1,1})^5 + \Gamma_{8} + \Gamma_8$---see e.g.~\cite{Ginsparg:1986bx,Font:2020rsk} for details of the isomorphism relevant to the relation between the $\GE_8\times\GE_8$ and $\Spin(32)/\Z_2$ heterotic strings---induces an isomorphism $\R^{d+16,d} \simeq R^{d,d}\oplus \R^8 \oplus \R^8$, and we can use this to pick a basis that isometrically respects this splitting.  For the $\R^{d,d}$ factor we choose lattice vectors $\be_i, \be^{\ast i}$ satisfying
\begin{align}
\be_i . \be^{\ast j} &= \delta^j_i~, &
\be_i . \be_j &= 0~, &
\be^{\ast i} . \be^{\ast j} &= 0~, &
\end{align}
while for each of the $\Gamma_8$ factors we choose the set of simple roots $\balpha_I$,  $I=1,\ldots,8$, of $\Le_8$, with $\balpha_I \cdot \balpha_J$ the Killing metric, taken to be negative in our conventions.  The $\balpha_I$ are encoded in the Dynkin diagram written in terms of the standard orthonormal basis for $\R^8$,  denoted by $\bv_a$,  with $a = 1,\ldots, 8$ and $\bv_a\cdot\bv_b = -\delta_{ab}$:
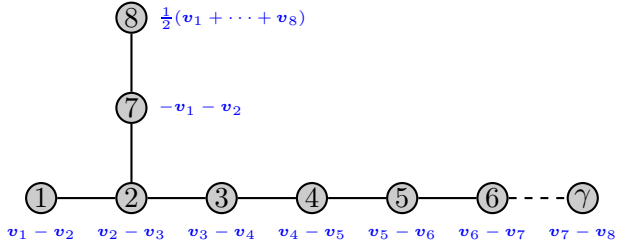
\begin{align*}
\begin{tikzpicture}
[root/.style={circle,draw, fill = black!20,thick,  inner sep = 0pt, minimum size = 4mm},scale=1.2]
\node (e1) at (1,0) [root] {$1$};
\node [blue,below] at (e1.south) {\tiny{$\bv_1 -\bv_2$}};
\node (e2) at (2,0) [root] {$2$};
\node [blue,below] at (e2.south) {\tiny{$\bv_2 -\bv_3$}};
\node (e3) at (3,0) [root] {$3$};
\node [blue,below] at (e3.south) {\tiny{$\bv_3 -\bv_4$}};
\node (e4) at (4,0) [root] {$4$};
\node [blue,below] at (e4.south) {\tiny{$\bv_4 -\bv_5$}};
\node (e5) at (5,0) [root] {$5$};
\node [blue,below] at (e5.south) {\tiny{$\bv_5 -\bv_6$}};
\node (e6) at (6,0) [root] {$6$};
\node [blue,below] at (e6.south) {\tiny{$\bv_6 -\bv_7$}};
\node (e7) at (2,1) [root] {$7$};
\node [blue,right] at (e7.east) {\tiny{$-\bv_1 -\bv_2$}};
\node (e8) at (2,2) [root] {$8$};
\node [blue,right] at (e8.east) {\tiny{$\ff{1}{2}(\bv_1+\cdots+\bv_8)$}};
\node (e0) at (7,0) [root] {$\gamma$};
\node [blue,below] at (e0.south) {\tiny{$\bv_7-\bv_8$}};
\draw[thick] (e1) -- (e2) -- (e3) -- (e4) -- (e5) -- (e6);
\draw[thick,dashed] (e6) -- (e0) ;
\draw[thick] (e2) -- (e7) -- (e8);
\end{tikzpicture}
\end{align*}
To accommodate the second (our hidden) $\Le_8$ factor, we will let the indices $I,a$ run through $9,\ldots,16$ as well, but this will play no role in our analysis.

With the lattice set up complete, a point in the Narain moduli space  corresponds to a choice of spacelike $d$-plane $\Pi$ spanned by the vectors $\bpi_i$, which can be taken to be of the form\footnote{We follow, with some small adjustments, the presentation given in~\cite{Aspinwall:1996mn}}
\begin{align}
\bpi_i & = \be_i + (g_{ij} + b_{ij} -\ff{1}{2} A_i \cdot A_j) \be^{\ast j} + A_i~, & A_i = A_i^I \balpha_I = A_i^a \bv_a~.
\end{align}
In addition to the torus metric $g_{ij}$, this encodes the choice of constant $B$-field $b_{ij}$ and Wilson lines $A_i^a$ valued the Cartan subalgebra of $\Le_8$.

The spacelike vectors satisfy $\bpi_i . \bpi_j = 2 g_{ij}$, and it is easy to find a basis for the orthogonal complement $\Pi_\perp$, a timelike $(d+16)$--plane.  We take it to be spanned by
\begin{align}
\bpit_i &= \bpi_i -2 g_{ij} \be^{\ast j} ~, &
\bpit_a & = \bv_a +A^{a}_i \be^{\ast i}~.
\end{align}
These two sets of vectors are orthogonal to the $\bpi_i$ and each other, and satisfy $\bpit_i .\bpit_j = -2g_{ij}$ and $\bpit_a .\bpit_b = -\delta_{ab}$.  Extending our discussion of the compact boson, we arrive at the vertex operators
\begin{align}
\cV_p & = ~:~ \bC(p) \exp\left[ \ff{i}{\sqrt{2}} (\bpi_i .p) \Phi^i_R(\zb) + \ff{i}{\sqrt{2}} (\bpit_i.p) \Phi^i_L(z) + i (\bpit_a .p) \cX_{L}^a (z) \right]~:~
\end{align}
for all 
\begin{align}
p &= w^i \be_i + n_i \be^{\ast i} + \lambda \in \Gamma_{16+d,d}~, &\lambda& \in \Gamma_8 +\Gamma_8~.
\end{align}
The weights of these operators are given by a generalization of~(\ref{eq:circleweights}):
\begin{align}
\label{eq:generalweights}
h_R(p) & = \ff{1}{4} (\bpi_i.p) g^{ij} (\bpi_j.p)~,&
h_L(p) & = \ff{1}{4} (\bpit_i.p) g^{ij} (\bpit_j.p) + \textstyle\sum_{a} \ff{1}{2} (\bpit_a.p)^2~,
\end{align}
and the spin satisfies
\begin{align}
s(p) = h_L(p) - h_R(p) & = -\ff{1}{2} p.p = -n^i w_i -\ff{1}{2}\lambda\cdot\lambda~.
\end{align}
Since $\Gamma_8$ is an even lattice the last term is an integer for all $\lambda$, so that $s(p) \in \Z$.  

Finally, to complete the heterotic construction we need to add the right-moving superpartners of the $\Phi^i$, the fermions $\psi^i(\zb)$.  For all of our applications we will be able to group these into the two Weyl fermions $\psi^{1,2}$, $\psib^{1,2}$ for the $\Ttor^4$ directions, and the extra Majorana-Weyl $\psi^5$ for the additional circle direction.

\subsection{Two orbifolds}
Starting with this presentation of the heterotic string on $\Ttor^5$, we now consider the special locus where the $\Ttor^4$ factor admits a $G=\Z_N$ symmetry with generator $g$.
The $\Z_N$ symmetry leads to significant simplification in the CFT moduli:  both $g_{i5}$ and $b_{i5}$ must be zero for $i\neq 5$, and the $A_{i\neq 5}$ may be set to zero as well.\footnote{Note the $A_{i\neq 5}$ Wilson lines are not required to be set to zero by our $\Z_N$ symmetry because there is an accompanying action on the left-moving fermions, or equivalently on the $\cX_a$.  In the orbifold theory these  symmetry-preserving Wilson line parameters describe hypermultiplet expectation values.}  

To understand how $G$ acts on the states, we reconsider the actions described in our RNS discussion.  These were the geometric orbifold action on $\Ttor^4$, accompanied by its supersymmetric extension to the right-moving fermions, the translation on $S^1$, and the action on the left-moving fermions $\gamma^{1,2}$.  We will now translate each of these into the bosonic description.

First, the geometric action on the $\Ttor^4$ bosons induces an action on the associated vectors $\vec{p} \in \Gamma_{d,d}$.  If the $\Phi^i$ coordinates transform as
\begin{align}
g_T \cdot \Phi^i = R^i_j \Phi^j~,
\end{align}
then
\begin{align}
\label{eq:gTaction}
g_T \cdot w^i &= R^i_j w^j~, &
g_T \cdot n_i & = (R^{-1})^j_i n_i~.
\end{align}
This preserves the spin of every $\cV_p$, and, because it is a torus isometry, it preserves the weights as well.  The right-moving fermions also transform accordingly, and as we discussed around equation~(\ref{eq:RNSZNaction}), it is convenient to represent the action by $g_{\psi} = \exp[\ff{2\pi i}{N} \Jb'_0]$.

Second, the translation in the circle direction is precisely the shift symmetry $g_{\text{sh}}$ described in section~\ref{ss:shiftorbifold}.  It leaves the lattice vectors invariant and acts on states by a $\Z_N$-valued phase.

Finally, we have the action on the $\gamma$ fermions, represented by $g_{\gamma} = \exp[\ff{2\pi i}{N} J'_0]$.  This is the only one that is not quite obvious in the bosonized description, but because we can write it in terms of the current $J'$, finding the corresponding action on the $\cX^a$ is not too difficult.  Taking a look at our Dynkin diagram, we see that
\begin{align}
J' & = i \p\cX^7 -i\p\cX^8~,
\end{align}
i.e. it is the current that corresponds to the extended root of the diagram.  Thus, $g_\gamma$ turns out to be another shift action in the bosonized description:
\begin{align}
g_\gamma |p\ra & = e^{\ff{2\pi i}{N} (\bv_7-\bv_8) . p} |p\ra~.
\end{align}
We can see this gives the expected structure for the unbroken symmetry with $A=0$.  In this case the $\Le_8$ currents that survive the projection are the Cartan currents $i\p \cX^a$, as well as the roots $\balpha_I$ orthogonal to $\bv_7-\bv_8$.  For $N>2$ this gives rise to the $\Le_7 \oplus \Lu(1)'$ current algebra, and in the special case of $N=2$ the state corresponding to the $\bv_7-\bv_8$ root is left invariant as well, leading to the $\Lu(1)'\to \su(2)'$ enhancement.

As above, we have two $\Z_N$ symmetries by which we can orbifold:  
$G$, generated by $g = g_T g_\psi g_\gamma g_{\text{sh}}$, leads to the $X_5$ orbifold, 
while $G'$, generated by $g' = g_T g_\psi g_\gamma$, produces the $\Ttor^4/\Z_N\times S^1$ orbifold.

Both of the symmetries are consistent with turning on a Wilson line $A_5$ along the shift circle, and by including this degree of freedom, we will demonstrate that, despite appearances, the two orbifolds are equivalent.  We will do this by finding an element $t$ of the T-duality group $\GO(\Gamma_{17,1}) \subset \GO(\Gamma_{21,5})$ such that 
\begin{align}
\label{eq:heteroticequivalence}
g'  = t^{-1} g t~.
\end{align}  
To give this construction, we will have to delve a little bit into the structure of the heterotic T-duality group.

\subsection{Elements of $\GO(\Gamma_{17,1})$}
The T-duality group $\GO(\Gamma_{16+d,d})$ arises from lattice isomorphisms, and each such isomorphism induces a (possibly trivial) action on the moduli.  A discussion of its generators is given in~\cite{Giveon:1994fu,Font:2020rsk}.  In our discussion two elements will play a role:  the T-duality transformation on the $5$-th circle, as well as a shift of the Wilson line associated to the circle by a lattice vector in $\Gamma_8$.  

Typically, the induced action on the moduli is complicated.  For example, a T-duality on a single circle usually involves a non-trivial action on the $g_{ij}$, $b_{ij}$ and $A_i$ components~\cite{Giveon:1994fu,Font:2020rsk}.    However, since we assume the $\Z_N$ symmetry, the moduli are restricted, so that we may as well think of our transformations as living in $\GO(\Gamma_{17,1})$.

A vector in $\Gamma_{17,1}$ has the form
\begin{align}
p & = w \be + n \be^\ast + \lambda~,
\end{align}
and using~(\ref{eq:generalweights}) with $g_{55} = r^2$, and $A_5 = A$, we find
the weight and spin
\begin{align}
h_R(p) & = \ff{1}{4r^2} \left(\pi.p\right)^2~, &
s(p) & = h_L(p) - h_R(p) =-\ff{1}{2} p.p =  -nw -\ff{1}{2} \lambda\cdot\lambda~,
\end{align}
with
\begin{align}
\pi.p & = n + \left(r^2 -\ff{1}{2} A\cdot A \right) w + A\cdot \lambda~.
\end{align}
The T-duality action on the lattice is simply the exchange $(n,w) \to (w,n)$.  Let us call this action $g_1(p)$.  Clearly $g_1(p).g_1(p) = p.p$ for all $p$, which is necessary and sufficient for $g_1$ to be a lattice isomorphism.  Moreover, we can find an induced action the moduli $(r,A)$, so that 
\begin{align}
(g_1^\ast h_R) (p) = h_R( g_1(p))~:
\end{align}
\begin{align}
(r',A') &= g_1^\ast(r,A)~, &
r' & = \frac{r}{r^2 -\ff{1}{2} A\cdot A}~,&
A' & = \frac{A}{r^2-\ff{1}{2} A\cdot A}~.
\end{align}
The lattice vector shift of the Wilson line arises through a more elaborate lattice isomorphism, depending on a choice of lattice vector $\rho \in \Gamma_8 + \Gamma_8$:
\begin{align}
g_2(p) = p + (\rho\cdot \lambda -\ff{1}{2} \rho\cdot\rho w) \be^\ast - \rho w~,
\end{align}
or more explicitly,
\begin{align}
g_2(n,w,\lambda) = (n +\rho\cdot \lambda -\ff{1}{2} \rho\cdot\rho w, w, \lambda-\rho w)~.
\end{align}
This action also preserves the lattice since $g_2(p).g_2(p) = p.p$~, and the induced map on the moduli is simply
\begin{align}
(r',A') & = g_2^\ast(r,A) = (r, A + \rho)~.
\end{align}
Having reviewed the form of these basic maps, we will now use them to construct an isomorphism between the two orbifold theories, i.e. the compactification on $X_5$ and the $\Ttor^4/Z_N \times S^1$ orbifold.

\subsection{Heterotic isomorphism}
Let $t = g_1 g_2$.  We claim that this combination of $\GO(\Gamma_{17,1})$ elements leads to the desired equivalence~(\ref{eq:heteroticequivalence}) for an appropriate choice of lattice vector $\rho$.  Since $t$ does not act on the $\Ttor^4$ bosons or right-moving fermions, and our two actions $g$ and $g'$ act in the same way on those degrees of freedom, it is sufficient to check the claim for $p \in \Gamma_{17,1}$.

We set the $\rho$ to be proportional to the root $\balpha_6 = \bv_6 - \bv_7$:
\begin{align}
\rho & = (1-N) \balpha_6~,
\end{align}
and we define for convenience $\bgamma = \bv_7-\bv_8$.  Note that $\bgamma\cdot \rho = 1-N$.  We now compute
\begin{align}
t |n,w,\lambda\ra & = | w, n+\rho\cdot\lambda-\ff{1}{2} \rho\cdot\rho w, \lambda-\rho w\ra~,
\end{align}
and therefore
\begin{align}
t^{-1} g t |n,w,\lambda\ra & = \exp\left[ \ff{2\pi i}{N} \left( w + \bgamma \cdot (\lambda-\rho w)\right) \right] |n,w,\lambda\ra = e^{\ff{2\pi i}{N} \bgamma \cdot\lambda} |n,w,\lambda\ra = g' |n,w,\lambda\ra~.
\end{align}
In the second equality the phase dependence on the winding mode $w$ drops out precisely because $\bgamma\cdot \rho = 1-N$.  Thus, despite appearances, the two orbifold CFTs are isomorphic!

\subsubsection*{Special features of $N=3$}
As we remarked above, the correspondence between the theories is particularly nice when $N=3$, since there it is possible to match the massless spectra by turning on a single Wilson line for the $\Lu(1)'$ vector multiplet.  We can see these special features from the point of view of our isomorphism as well.  Precisely when $N=3$ we can set the Willson line shift to be $\rho = \bgamma$, and the winding mode will again drop out from the phase factor.    

In this case, then, we can identify the $X_5$ theory with radius $r$ and circle Wilson line $A_5 = a \bgamma$ with the $\Ttor^4/\Z_3 \times S^1$ theory with radius  $r'$ and circle Wilson line $A_5 = a'\bgamma$ via
\begin{align}
r &= \ff{r'}{r'^2+(a'-1)^2}~,&
a &= \ff{a'-1}{r'^2+(a'-1)^2}~.
\end{align}
As expected, the duality is a stringy one.  For example, setting $a'=1$, which corresponds to the $a=0$ $X_5$ theory analyzed above, we find that a large radius circle on the $\Ttor^4/\Z_3 \times S^1$ side corresponds to $X_5$ with a small shift circle.

Note that the theory with $a' = 1$ is equivalent to $a'=0$ in the $\Ttor^5$ theory, but that is not the case in the orbifold, where instead the equivalence is $a' \sim a' + 3$.  This is closely related to the appearance of the fractional $\GU(1)'_L$ charges in the twisted sector states, and it is thus not surprising that such massless states are lifted for $a'=1$.

The $a'=0$ locus is also interesting, since it corresponds to the $\Ttor^4/\Z_3\times S^1$ theory with the massless hypermultiplets coming from the twisted sectors.  It would be surprising if we could identify it in the $X_5$ description at large radius, and indeed this is not so.  When written in terms of the $X_5$ moduli, the $a'=0$ locus corresponds to the semicircle with $r\ge 0$ and 
\begin{align}
(2r)^2 + (1+2a)^2 = 1~,
\end{align}
so that $r \le 1/2$.

\subsection{Cocycle subtleties and their resolution} \label{ss:cocycles}
The reader who has made it this far may wonder if we have not introduced more formalism than we need to describe our results:  we have not used the structure of the vertex operators $\cV_p$ in any detail, and the operators $\bC(p)$, which are typically relegated to a last section or appendix have played no role in our discussion.  We will now explain the need for these terms and point out that in general their presence can lead to subtleties in the analysis of CFT symmetries.  We will also see, however, that in this case we are lucky:  all of the potential subtle factors drop out, and the duality conclusions reached in the previous section remain unmodified.  Nevertheless, since the general observations here may lead to subtleties in closely related discussions, we will include them.  In addition to the standard textbook references~\cite{Green:1987mn,Polchinski:1998rq}, which supply some of the background, our thinking about these issues was guided by the work~\cite{Tan:2015nja}.

\subsubsection*{A class of CFT symmetries}
We are interested in discussing a class of CFT symmetries of Narain compactification that are realized as the following $G$ action on the vertex operators.  For every $g\in G$ there is a map $\varphi_g : \Gamma\to \Gamma$ and a factor $U(g,p) \in \C$ such that
\begin{align}
g\circ \cV_{p} = U(g,p) \cV_{\varphi_g(p)}~.
\end{align}
Not all CFT symmetries can be realized in this fashion.  For example, the $\SU(2)\times\SU(2)$ symmetries of the compact boson at self-dual radius take a more general form and also mix the $\cV_p$ with non-Kac-Moody primary operators such as $\p\Phi$ and $\pb\Phi$.  

We wish the $g$-action to be invertible and to be consistent with the OPE, and it must preserve the weights of the operators.  Therefore it must be that for every $g$ the factor $U(g,p) \in \C^\ast$, and $\varphi_g$ is a lattice isomorphism.  We will insist that the action is unitary, which means $U(g,p)$ is a pure phase.  Since we also want composition to be consistent with the group product structure, i.e.
\begin{align}
(g_2 g_1) \circ \cV_{p} = g_2 \circ \left( g_1\circ \cV_p\right)~,
\end{align}
We also learn that the phases must obey
\begin{align}
U(g_2 g_1,p) = U(g_2, g_1(p)) U(g_1,p)~,
\end{align}
and similarly the $\varphi_{g}$ should satisfy $\varphi_{g_1g_2}(p) = \varphi_{g_1}(\varphi_{g_2}(p))$.  We now see that $\varphi$ must be a map to the group of automorphisms of the lattice, i.e. $\varphi:  G\to \GO(\Gamma)$.  The resulting subgroup of $\GO(\Gamma)$
\begin{align}
G_{\Gamma} \simeq G/\ker(\varphi)
\end{align}
is in general smaller than $G$: for example, the circle shift symmetry has $G =\Z_N$ and $G_\Gamma = 1$.

These are sensible constraints determined by the group structure, but there are further constraints on the factors $U(g,p)$, and this is where the cocycles make an appearance.

\subsubsection*{A look at the cocycles}
The factors $\bC(p)$ are introduced to resolve an issue with commutation properties of the naive vertex operators $\cV_p^{\text{naive}}$ which lack these factors:  the OPE of two such operators has a non-trivial monodromy as we transport one operator around the other, which leads to 
\begin{align}
: \cV^{\text{naive}}_{p_1} (-z/2 ) :~ : \cV^{\text{naive}}_{p_2} (z/2) : ~= 
e^{i\pi p_1.p_2} : \cV^{\text{naive}}_{p_2} (z/2) :~: \cV^{\text{naive}}_{p_1} (-z/2 ) :~.
\end{align}
Thus, the operators appear to anticommute whenever $p_1.p_2$ is odd.  To resolve this, the operators are modified to include the $\bC(p)$, which are chosen so that
\begin{align}
\label{eq:cocyclephase}
\cV_q \bC_p  & =\varepsilon(q,p) \bC_p {\cV_q}~, & \bC_p \bC_q &= \bC_{p+q}~
\end{align}
for some phase $\varepsilon(q,p)$.  Including these factors will remove the unwanted factor $e^{i\pi p_1.p_2}$ if the phases obey
\begin{align}
\label{eq:correctcommutator}
\varepsilon(p_2,p_1) = (-1)^{p_1. p_2} \varepsilon(p_1,p_2)~.
\end{align}
Associativity of the OPE places further non-trivial conditions on the phases:
\begin{align}
\label{eq:cocyclelinearity}
\varepsilon(p_1,p_3) \varepsilon(p_2,p_3) &= \varepsilon(p_1+p_2,p_3)~,
\end{align}
and
\begin{align}
\label{eq:cocycle}
\varepsilon(p_1,p_2) \varepsilon(p_1+p_2,p_3) =\varepsilon(p_1,p_2+p_3)\varepsilon(p_2,p_3) ~.
\end{align}
The second of these justifies the name of cocycle, i.e. $\varepsilon\in H^2(\Gamma,\GU(1))$.  Consider the group homomorphisms $\Gamma\times\Gamma \to \GU(1)$, denoted by $c(p_1,p_2)$.  The coboundary map to the group of homomorphisms $\Gamma\times\Gamma\times\Gamma\to\GU(1)$ is taken to be
\begin{align}
\delta_{32} c(p_1,p_2,p_3) & = \frac{c(p_2,p_3)}{c(p_1+p_2,p_3)} \frac{c(p_1,p_2+p_3)}{c(p_1,p_2)}~,
\end{align}
Similarly, starting with homomorphisms $\Gamma\to\GU(1)$ denoted by $f(p)$, we have
\begin{align}
\delta_{21} f(p_1,p_2) & = \frac{ f(p_1) f(p_2)}{f(p_1+p_2)}~,
\end{align}
and the reader can check $\delta_{32} (\delta_{21} f) =1$.  The condition~(\ref{eq:cocycle}) is  the statement that $\delta_{32}\varepsilon =1$, i.e. that $\varepsilon$ is a cocycle that defines a class in $H^2(\Gamma,\GU(1))$.  On the other hand, the condition~(\ref{eq:correctcommutator}) shows that $\varepsilon(p_1,p_2)$ cannot be a coboundary.

As shown in the references~\cite{Green:1987mn,Polchinski:1998rq,Tan:2015nja}, it is possible to choose the $\bC(p)$ so that $\varepsilon$ take the following form:
\begin{align}
\varepsilon(p_1,p_2) & = \exp\left[\ff{i\pi}{2} (n_{1i} w_2^i - w_1^i n_{2i}) + i\pi\Omega(\lambda_1,\lambda_2)\right]~,
\end{align}
where the bilinear antisymmetric form $\Omega(\lambda_1,\lambda_2)$ is obtained by writing $\lambda_{1,2} = \lambda_{1,2}^I \balpha_I$ in terms of the simple roots and then setting
\begin{align}
\Omega(\lambda_1,\lambda_2) & = \ff{1}{2} \sum_{I>J} \left(\lambda_1^I\lambda_2^J-\lambda_2^I\lambda_1^J\right) \balpha_I \cdot\balpha_J~.
\end{align}

\subsubsection*{Cocycles and symmetry phases}
Finally we come to our key point:  the cocycle factors and the phases $U(g,p)$ are intertwined, and the consistency of the OPE with the symmetry action, i.e.
\begin{align}
~:g \circ \cV_{p} (z_1):~ : g \circ \cV_{q}(z_2):~ =~ g \circ \left(~ :\cV_{p}(z_1):~: \cV_{q}(z_2):~ \right)~
\end{align}
require the phase factors to obey~\cite{Tan:2015nja}
\begin{align}
\label{eq:factorsfromcocycles}
\frac{U(g,p+q)}{U(g,p) U(g,q)} & = \frac{\varepsilon(\varphi_{g} (p),\varphi_{g} (q))}{\varepsilon(p,q)}~.
\end{align}
There are two ways to read this equation.  On one hand, it shows that the ratio of cocycles $\frac{\varepsilon(\varphi_{g} (p),\varphi_{g} (q))}{\varepsilon(p,q)}$ is a coboundary, and $U(g,p)$ is the trivializing cycle.  On the other hand, whenever the ratio is not $1$, it gives an obstruction to choosing $U(g,p)$ to be a homomorphism from the lattice to $\GU(1)$.  

There are two obvious questions about this result:  how does the it depend on the choice of cocycle?  to what extent does it determine the phases $U(g,p)$?  We leave it to the reader to check that a modification of $\varepsilon$ by a coboundary $f(p)$ modifies $U(g,p) \to U(g,p) f(g(p))/f(p)$, and the new $U(g,p)$ so obtained is consistent with the group structure.  As far as the second question goes, it is clear that if $U_1(g,p)$ and $U_2(g,p)$ both solve~(\ref{eq:factorsfromcocycles}) with the same $\varepsilon$, then their ratio $U_1/U_2$ is a homomorphism from $\Gamma$ to $\GU(1)$ for all $g$.

\subsubsection*{Some examples}
Let us consider some simple examples that illustrate the appearance (and non-appearance) of non-trivial phase factors $U(g,p)$.  By non-trivial we mean here factors that are not in $\Hom(\Gamma,\GU(1))$ and whose appearance is necessitated by~(\ref{eq:factorsfromcocycles}).

By far the simplest example is any action where $\varphi_g(p) = p$ for all $p$.  In this case, $U(g,p)$ must be in $\Hom(\Gamma,\GU(1))$.  The circle shift orbifold yields a simple example.  A bit more interesting is any geometric symmetry that arises from a $\GL(d,\Z)$ change of coordinates on the torus, such as our symmetry $g_T$ in~(\ref{eq:gTaction}), acting on the winding and momentum modes as in~(\ref{eq:gTaction}).  However, in this case too we find $\varepsilon(\varphi_g(p),\varphi_g(q))  = \varepsilon(p,q)$, so that again the phase can be taken to be trivial.

For a more interesting example in $\Gamma_{d,d}$ we can consider integral shifts of the $b$-field background, for which
\begin{align}
\varphi_g (n_i,w^i) = (n_i + \Theta_{ij} w^j, w^i)~,
\end{align}
where $\Theta_{ij}$ is an arbitrary integral antisymmetric matrix.  In this case
\begin{align}
\frac{\varepsilon(g(p_1),g(p_2)) }{\varepsilon(p_1,p_2)} =  \exp\left[i \pi \textstyle\sum_{i>j}\Theta_{ij} (w_1^i w_2^j - w_2^iw_1^j)\right] \neq 1~,
\end{align}
and a non-trivial phase is required to satisfy~(\ref{eq:factorsfromcocycles}).  A choice that works is to take
\begin{align}
U(g,p) & = \exp\left[i \pi \textstyle\sum_{i>j}\Theta_{ij} w^i w^j \right]~.
\end{align}
Since the $g$--action preserves winding numbers $U(g,g(p)) = U(g,p)$, so that $U(g^k,p) = U(g,p)^k$, which is consistent with the group structure.

For a more elaborate example, we consider the Wilson line shift in $\Gamma_{17,1}$ relevant to our construction:
\begin{align}
\varphi_g(n,w,\tau) = (n + \rho\cdot\lambda - \ff{1}{2} \rho\cdot\rho w, w,\lambda-\rho w)~.
\end{align}
In this case a somewhat lengthy computation, making creative use of mod $2$ conditions in the exponential, leads to
\begin{align}
\frac{\varepsilon(\varphi_g(p_1),\varphi_g(p_2))}{\varepsilon(p_1,p_2)} =
\exp\left[ i \pi \left( w_1 S(\rho,\lambda_2) + w_2 S(\rho,\lambda_1) \right)\right]~,
\end{align}
where the bilinear map $S$ is 
\begin{align}
S(\rho ,\lambda) = \sum_I \rho^I \lambda^I - \sum_{I>J} \balpha_I \cdot \balpha_J\rho^I  \lambda^J \in \Z~. 
\end{align}
The naive guess $U(g,p) = e^{i\pi w S(\rho,\lambda)}$, while solving~(\ref{eq:factorsfromcocycles}), does not satisfy the group property
\begin{align}
U(g^2,p) = U(g,g(p)) U(g,p)~.
\end{align}
We can fix this by multiplying the naive guess by an element of $\Hom(\Gamma,\GU(1))$.  We find that
\begin{align}
U (g,p) & = \exp\left[i\pi w S(\rho,\lambda + \rho/2)\right]~
\end{align}
satisfies all of the desired properties.

\subsubsection*{Duality unmodified}
We just discussed at length that symmetry actions of the sort we used to prove the duality between $X_5$ and $\Ttor^4/\Z_N\times S^1$ heterotic compactifications in general come with extra phase factors.  Since the whole point of our duality claim was a match of phase factors for two group actions, this is a non-trivial concern.

To allay this worry, we write the correct form for the symmetry action on the states, taking into account the possible extra phases.  In keeping with the spirit of this section, we distinguish in our notation between the action on the states and the action on the lattice vectors, so that we have, coming back to~(\ref{eq:heteroticequivalence}),
\begin{align}
t^{-1} g t |p\ra & = U(t,p) U(g,\varphi_t(p)) U(t^{-1} ,\varphi_{gt}(p)) |\varphi_{t^{-1} g t} (p) \ra~.
\end{align}
The group multiplication properties imply the relation
\begin{align}
1 = U(t^{-1}t, \varphi_{t^{-1}} (p) ) = U(t^{-1},p) U(t,\varphi_{t}^{-1}(p))~,
\end{align}
so that 
\begin{align}
U(t^{-1},p) = \frac{1}{U(t,\varphi_{t}^{-1}(p))}~.
\end{align}
But, since we have $\varphi_{t^{-1} g t} (p) = p$, the phase factor becomes
\begin{align}
t^{-1} g t |p\ra & = U(t,p) U(g,\varphi_t(p)) \frac{1}{U(t,p)} |\varphi_{t^{-1} g t} (p) \ra = U(g,\varphi_t(p)) |p\ra~,
\end{align}
but that is exactly the factor we showed to be equal to $U(g',p)$.

We neglected two points in this discussion: the phases associated to the right-moving fermions, and the $\Z_N$ action on $\Ttor^4$.  As we saw above, the latter does not lead to any additional phase factors, and the former is identical in both theories.

\section{Gerbes and duality in type II compactification}
We have seen so far that in the context of the heterotic string T-duality can relate compactifications on $X_5$ to more familiar compactifications on $\Ttor^4/\Z_N \times S^1$, but the relationship cannot be seen at the level of the NS sector and $\GO(d,d,\Z)$ transformations---the Wilson line background is essential.  It is therefore not surprising that a similar construction is unavailable in type II compactifications: there is no transformation $t \in \GO(d,d,\Z)$ such that~(\ref{eq:heteroticequivalence}) holds.  

Nevertheless, we can still ask:  what is the T-dual of $X_5$?  We will argue that the dual geometry is indeed that of $\Ttor^4/\Z_N\times S^1$, but this geometry is equipped with a choice of a flat $B$-field, which leads to drastic modifications in the spectrum of the theory.  

Some of the backgrounds that we discuss involve $B$-field data that is pure holonomy:  the connection $B= 0$, but the gerbe background is nevertheless topologically non-trivial.  To describe these global structures we will need to introduce some machinery of abelian gerbes, and in our context these arise as $G$-equivariant topologically trivial gerbes on $\Ttor^5$.

It is not hard to see why such structures should arise naturally in our context.  We are interested in geometries that are presented as quotients $\Ttor^5/G$ and that also have a presentation as a circle fibration
\begin{equation*}
\begin{tikzcd}
S^1 \arrow[r,hook] & X \arrow[d] \\ & \Ttor^4/G
\end{tikzcd}~.
\end{equation*}
In each case the fibration is flat, meaning that the transition functions can be taken to be constant, and the connection set identically to zero.  Nevertheless, the fibration is topologically non-trivial, with holonomy encoding the topological data.  On the other hand, T-duality in general exchanges the topology of the circle fibration for that of the $B$-field~\cite{Bouwknegt:2003vb}, and we might expect this to be the case in our context as well.   We will develop this picture, but to motivate some of the developments, we will make a small digression on T-duality with $H$-flux, following~\cite{Bouwknegt:2003vb}.

\subsubsection*{Correspondence spaces and vertical gerbes}
Suppose $X$ is the total space of a principal circle bundle $\pi_g : P_g \to M$.  This space has a nowhere vanishing vector field associated to the circle fiber, which we denote $V$ and a dual form $\Theta_g$ that satisfies
\begin{align}
d\Theta_g = \pi^\ast_g(F_g)~,
\end{align} 
where $F_g$ is the curvature of the circle bundle.  An $H$-flux over $X$ is said to be T-dualizable if $d H = 0$ and $V \llcorner H = \pi^\ast(F_b)$ for some closed $2$-form $F_b$.  This form is necessarily integral and defines (up to torsion) another principal circle bundle $\pi_b : P_b \to M$ with $1$-form $\Theta_b$ satisfying $d\Theta_b = \pi^\ast_b(F_b)$.

When this is the case, $H$ decomposes as
\begin{align}
H = \Theta_g \wedge \pi^\ast_g(F_b) - \pi^\ast_g (\Omega)~,
\end{align}
where $\Omega$ is a $3$-form on $M$ that satisfies $d\Omega = F_g \wedge F_b$, and we can also define the T-dual $H$-flux on the total space of $P_b$ by 
\begin{align}
\Hh = \Theta_b \wedge \pi^\ast_b(F_g) - \pi^\ast_b(\Omega)~.
\end{align}
This $H$-flux is clearly T-dualizable, and it can be argued that the  exchange of fibrations and fluxes implements the action of T-duality on various generalized (in general twisted) cohomology theories such as K-theory, as well as reproducing the Buscher rules~\cite{Bouwknegt:2003vb}.\footnote{The reference \cite{Bouwknegt:2003vb} treats the general case of a principal torus fibration, but for us the circle fibration will be sufficient.} 

The correspondence space is the geometric setting for these actions: it is the space  $Z = P_g \times_{M} P_b$, which fits into the commutative diagram
\begin{equation}
\begin{tikzcd}
~ & Z \ar[ld, "p_g"] \ar[rd,"p_b"'] &~  \\
P_g  \ar[rd, "\pi_g"]&~ & P_b \ar[ld,"\pi_b"'] \\
~  & M & ~
\end{tikzcd}
\end{equation}
The pullbacks of the two fluxes $p_g^\ast(H)$ and $p_b^\ast(\Hh)$ differ by an exact form: $d \left( p_b^\ast(\Theta_b) \wedge p_g^\ast(\Theta_g)\right)$.

Turning the construction around, given two principal circle bundles over $M$ with curvatures $F_g$ and $F_b$ satisfying $[F_g \wedge F_b] = 0 \in H^4(M,\R)$, we can choose $H$-fluxes on each of them that will be exchanged by T-duality.

A special case of the construction that is of particular importance for us is to take $H = \Theta_g \wedge \pi^\ast_g(F_b)$---this is the analogue of what we will call a ``vertical'' gerbe.  While elegant, this comes at a heavy price:  the curvature forms have to satisfy $F_g \wedge F_b = 0$ pointwise, and not merely in cohomology.

Ordinarily the discussion of string backgrounds with $H$--flux is complicated by basic questions about existence of a such a solution in string theory.  Our constructions will not have this difficulty since $H = 0$, and the correspondence space will give an elegant encoding of T-dual theories with flat principal circle bundles.  But we will also see torsional echoes of the $F_g \wedge F_b = 0$ constraint:  not every pair of circle bundles over $M = \Ttor^4/G$ can be a correspondence space.  To describe those results, we will now delve into equivariance.

\subsection{Equivariant flat line bundles}

Let $X$ be a compact smooth manifold with an action by a discrete group $G$, i.e. there are diffeomorphisms $\varphi_a: X \to X$ obeying $\varphi_{a} (\varphi_{b}(x)) = \varphi_{ab} (x)$ for all $a,b\in G$.   We are interested in describing flat line bundles over the quotient $M=X/G$, and this can accomplished by describing $G$--equivariant line bundles on $L \to X$.  

Recall that a line bundle is flat if and only if its curvature is zero.  In this case it is possible to choose a trivialization such that the transition functions are constant, and the connection is identically zero.  Since line bundles are topologically classified by $c_1(L) \in H^2(X,\Z)$, the first Chern class of a flat line bundle is a torsion class.  When $H^2(X,\Z)$ (or, equivalently, $H_1(X,\Z)$) is torsion-free, then every flat line bundle is trivial.

A  line bundle $\pi : L \to X$ is $G$-equivariant if there is a lift of the $G$-action on $X$ to a $G$-action on $L$ compatible with the projection $\pi$, where $G$ acts linearly on the fiber directions.  The quotient $L/G$ is then naturally a line bundle over $M = X/G$.  While in general a bundle may not admit a lift, every trivial bundle does admit a lift. In our setting, where $X = \Ttor^d$, every flat line bundle is trivial:  we can always find a lift of the $G$ action, and the characterization of flat line bundles over $M$ is then equivalent to the characterization of these lifts.\footnote{This situation is not special to the torus:  for any $M$ we can set $X$ to be its universal cover, with $G$ the group of deck transformations; in this case we find the explicit characterization of flat line bundles on $M$ via representations $\rho:  H_1(M,\Z) \to \GU(1)$\cite{Kobayashi:1987VB}.}

The reader may wish to consult the appendix~\ref{ss:cechlinebundle} for a review of some general aspects of such constructions, but in this section we will stick to the concrete setting relevant for our examples:  $G = \Z_N$,   $X = \Ttor^d$, and $L = X\times \C$ is the trivial line bundle with trivial transition functions and a torus-invariant connection $A$.  

We lift the $G$ action on $X$ to a $G$ action on the total space $L$ by specifying circle-valued functions $R^a: X\to S^1$ that obey\footnote{As described in more detail in the appendix~\ref{ss:cechlinebundle}, for a general line bundle $L\to X$ the $R^a$ need not be globally defined circle-valued functions, but when $L$ is trivial, the $R^a$ can be taken to be globally defined.}
\begin{align}
\label{eq:Rcomposition2}
R^{ba} = R^a\varphi_a^\ast(R^b)~.
\end{align}
Note that this composition is automatically associative:  $R^{(cb)a} = R^{c(ba)}$.

The $G$-action on a point $(x,\xi) \in X\times \C$ is 
\begin{align}
\psi &: L \to L~, & 
\psi (x, \xi) =(\varphi_a(x), R^a(x) \xi)~.
\end{align}
We also demand that the $G$ action preserves the connection, which  requires\footnote{We take our connections to be real.}
\begin{align}
\label{eq:equivariantconnection}
\varphi_a^\ast (A) = A -i d\log R^a~.
\end{align}
We assume that $A$ is a torus-invariant $1$-form on $\Ttor^d$, and so we will naturally demand that $d\log R^a $ is torus-invariant as well.

When $G= \Z_N$ and $X = \Ttor^d$, we can describe the $R^a$ explicitly.  Think of $\Ttor^d = \R^d/\Z^d$ with coordinates $\vec{x} \sim \vec{x} + \vec{m}$, where $\vec{m} \in \Z^d$.  Let $a$ be the generator of $\Z_N$ which acts by a linear transformation $a:  \vec{x} \to \varphi_a (\vec{x})$ on the torus coordinates, and denote by $R^n$ the function corresponding to the element $a^k$, defined by
\begin{align}
\label{eq:RdefZN}
R^1 & = e^{2\pi i \left( \ff{k}{N} + \vec{k} \cdot \vec{x}\right) }~,&
R^n & = R^1 \varphi_a^\ast \left( R^{n-1} \right)~, \quad k = 2,\ldots, N-1~.
\end{align}
These are circle-valued functions for $\vec{k} \in \Z^d$, and
when $k \in \Z$ we find $R^N = 1$, so that this indeed gives a representation of $\Z_N$. 

There is a further equivalence on these lifts, because given a circle-valued function $h: X\to S^1$ we can make a global gauge transformation which acts by
\begin{align}
R^1 \to R^1 \frac{\varphi_a^\ast(h)}{h}~,
\end{align}
and choosing $h = e^{2\pi i \vec{r} \cdot \vec{x}}$, we find an equivalence $\vec{k} \sim \vec{k} + M \vec{r}$, where 
\begin{align}
\label{eq:Mmatrix}
M = (\varphi_a^T - \iden)~.
\end{align}  
A constant equivariant connection that is consistent with this action is
\begin{align}
A_{\vec{k}} = 2\pi (M^{-1} \vec{k}) \cdot d\vec{x}~.
\end{align}

Thus, the gauge-inequivalent lifts of the $\Z_N$ action on $\Ttor^d$ are characterized by $k \in \{0,1,\ldots,N-1\}$ and $\vec{k} \in \Z^d /\im M$.  It is not hard to show that these are in 1:1 correspondence with representations $\rho : H_1(\Ttor^d/G, \Z) \to \GU(1)$, which topologically characterize flat line bundles on the quotient $M = \Ttor^d/G$~\cite{Kobayashi:1987VB}, and this shows that the line bundles on the quotient are in general non-trivial:  the first Chern class, while torsion, is non-zero.

The total space of the associated circle bundle  $P_g \to M$ is then also easy to understand:  it is the quotient of $\Ttor^{d+1}$ by the $\Z_N$ action generated by
\begin{align}
\label{eq:ourZN}
a(\vec{x}, x_{d+1}) = \left(\varphi_a(\vec{x}), x_{d+1} + \ff{k}{N} + \vec{k} \cdot \vec{x}\right)~,
\end{align}
with
\begin{align}
k &\in \{0,1,\ldots, N-1\}~, &
\vec{k} & \in \Z^d/\im M~.
\end{align}
Note that the compact flat manifolds described above arise naturally in this setting:  taking $d=4$ and setting $\vec{k} = 0$ and $k=1$, we obtain exactly the compact flat manifolds $X_5$.  More generally, we will denote these spaces by $X_{k,\vec{k}}$~.  We also note that this same classification of equivariant line bundles was obtained in~\cite{deBoer:2001wca} by calculating equivariant cohomology groups.

\subsubsection*{(Orbifold) diffeomorphism types of $X_{k,\vec{k}}$}
Although each choice of $k$ and $\vec{k}$ as above yields distinct circle bundles over $\Ttor^4/\Z_N$, Not all possible choices of $k,\vec{k}$ give distinct geometries for the total space.\footnote{We say that two orbifolds $X/G_1$ and $X/G_2$ are diffeomorphic if their $G$--actions are related by a diffeomorphism of $X$.}

As an example, consider the $\Z_2$ quotient, where $\varphi_a(\vec{x}) = -\vec{x}$,  and we can take the parameters $k \in \{0,1\}$, and $k_i \in \{0,1\}$.  To characterize the inequivalent orbifolds we can make changes of coordinates on $\Ttor^5$ as follows.

Suppose $\vec{k} \neq 0$.  Then, because $k_i \in \{0,1\}$, there is a $\SL(4,\Z)$ change of coordinates $\vec{y} = A \vec{x}$, such that $y_4 = \vec{k} \cdot\vec{x}$.   So, without loss of generality we set $\vec{k} \cdot \vec{x} = x_4$.  Now define a new coordinate  $y_4 = x_4 +\ff{k}{2}$.  Since $a(y_4) = -y_4 + 2k \sim -y_4$, the quotient is equivalent to the action
\begin{align}
a (\vec{x},x_5) = (-\vec{x}, x_5 +x_4)~.
\end{align}
Combining this with the remaining possibilities, we obtain $3$ diffeomorphism classes for the $\Z_2$ quotients, represented by
\begin{align}
\label{eq:Z2geometries}
\text{orbifold} && \text{quotient action} \nonumber \\[2mm]
X_{0,0} = \Ttor^4/\Z_2 \times S^1 &&  (-\vec{x}, x_5)  \nonumber\\
X_{1,0}~ \text{(smooth)} &&  (-\vec{x}, x_5+\ff{1}{2})  \nonumber\\
X_{0,\vec{k}} &&  (-\vec{x}, x_5+x_4)
\end{align}
A similar treatment can be given for the remaining $\Z_N$ actions, with similar results.  We relegate the full list to the appendix and here just summarize the key features.
\begin{enumerate}
\item For $N=3,4$ there are the analogous $3$ classes to the ones we found for $\Z_2$, with the replacement $\ff{1}{2} \to \ff{1}{N}$.
\item  For $\Z_4$ there is an additional class $X_{2,0}$.
\item For $\Z_6$ it is possible to set $ \vec{k} = 0$ by a change of coordinates, and there are just four inequivalent quotients with $0\le k \le 3$:
\begin{align}
\label{eq:Z6quotients}
\text{orbifold} && \text{quotient action} \nonumber \\[2mm] 
X_{k,0} &&  (\varphi_a(\vec{x}), x_5+\ff{k}{6})~.
\end{align}
\end{enumerate}
These results are consistent with the orbifold classification~\cite{Fischer:2012qj}.

\subsubsection*{Invariant form associated to circle isometry}
As we saw above, the discussion of T-duality on a circle bundle $P_g$ involves the nowhere vanishing $1$-form $\Theta$.  In our construction a natural guess is to set $\Theta = dx_{d+1}$, but when $\vec{k} \neq 0$ the differential $dx_{d+1}$ is not $G$--invariant. We can compensate for its transformation by using the connection~$A_{\vec{k}}$:
\begin{align}
\label{eq:Theta}
\Theta = dx_{d+1} - \ff{1}{2\pi} A_{\vec{k}}~.
\end{align}
$\Theta$ is not necessarily an integral form on $\Ttor^{d+1}$.  Denoting a homology cycle $\gamma \in H_1(\Ttor^{d+1},\Z)$ by $(\vec{m},m)$, we find
\begin{align}
\int_{\gamma} \Theta_{d+1} = m -\vec{m} \cdot M^{-1} \cdot \vec{k}~.
\end{align}

\subsection{Vertical equivariant flat gerbes}
Just as $G$--equivariant line bundles on $X$ can be used to describe line bundles on the quotient $M = X/G$,  $G$--equivariant gerbes on $X$ describe gerbes on $M$, and the data that specifies the lift of the $G$--action to the gerbe enters the computation of the partition function for the orbifold CFT associated to the quotient~\cite{Sharpe:2000ki,Sharpe:2003cs}.
We provide a detailed view of the construction in appendix~\ref{ss:appgerbes}, while in this section we describe the gerbes on $X$ and the $G$--action relevant to our examples.

In a nutshell, the $G$--action on a gerbe over $X$ is encoded in a set of gerbe gauge transformations, one for each element $a \in G$.  When $H^2(X,\Z)$ is torsion-free and the gerbe is topologically trivial and flat, meaning $H = dB = 0$, the relevant gerbe gauge transformations reduce to a choice of line bundles $\cL^a \to X$ equipped with connections $\cA^a$ such that $ \cL^b \otimes\varphi_b^\ast(\cL^a) \simeq\cL^{ab}$, and 
\begin{align}
\label{eq:babyBA}
\varphi_a^\ast(B) &= B + d\cA^a~, &
\cA^{ab} & = \cA^b + \varphi_b^\ast(\cA^a) + i d\log h^{a,b}~.
\end{align}
The $h^{a,b}$ are gauge transformations that realize the bundle isomorphism, and associativity of the group product requires these to satisfy
\begin{align}
\label{eq:hconsistency}
\varphi_c^\ast(h^{a,b} ) h^{ab,c} = h^{a,bc} h^{b,c}~.
\end{align}
Given any solution to these constraints and a representative $\lambda^{a,b}$ of a class in $H^2(G,\GU(1))$, we can obtain another solution by setting $h^{a,b}_{\text{new}} = \lambda^{a,b} h^{a,b}$.  On the other hand, multiplying $h^{a,b}$ by a group coboundary can be absorbed into gauge transformations on each of the bundles $\cL^{a}$ and gives the same $G$--action on the gerbe.  This choice of $[\lambda] \in H^2(G,\GU(1))$ is
the discrete torsion of~\cite{Vafa:1986wx}.  Since $H^2(\Z_N,\GU(1)) = 1$ this degree of freedom is absent in all of our constructions.  However, as emphasized in~\cite{Sharpe:2000ki},  $G$--equivariance allows for additional topological data associated to the holonomy of the gerbe.

We are interested in a particular class of gerbes defined by the following data:  a manifold $Y = X \times S^1$ equipped with a closed $1$-form $\Theta$ and a flat line bundle that is the pull-back of a bundle $L \to X$ with flat connection $\Lambda$, which is itself a global $1$-form on $X$.\footnote{Recall that every flat line bundle admits a trivialization with constant transition functions, and the connections compatible with this choice are global $1$-forms on $X$}  In this case, we obtain a flat ``vertical'' gerbe by setting
\begin{align}
B = \Lambda \wedge \Theta~.
\end{align}
Now suppose further that there is a $G$ action on $Y$ such that $\Theta$ is $G$--invariant, while $L$ is $G$--equivariant with the lift described by a choice of  phases $\Rt^a$.  In this case the vertical gerbe may admit a $G$--action as well.  To explore when this is the case, we simply compute
\begin{align}
\varphi_a^\ast (B) - B = d \left( - i \log(\Rt^a)  \Theta\right)~,
\end{align}
so that a naive solution to~(\ref{eq:babyBA}) is to set
\begin{align}
\cA^a = - i \log(\Rt^a) \Theta~.
\end{align}
There are three reasons why this construction is naive.  First, the $\cA^a$ so defined may not be a connection on a line bundle $\cL^a$ because $\ff{1}{2\pi} d\cA^a$ is not properly quantized.  Second, it may not be possible to find $h^{a,b}$ to satisfy~~(\ref{eq:babyBA}) and~(\ref{eq:hconsistency}). 

Turning to our specific interest with $Y = \Ttor^{d+1}$, $G = \Z_N$ acting as in~(\ref{eq:ourZN}), and $\Theta$ as in~(\ref{eq:Theta}), we set---$\beta$ and $\vec{\beta}$ are parameters analogous to $k$ and $\vec{k}$---
\begin{align}
\Rt^{1} &= e^{2\pi i \left( \ff{\beta}{N} + \vec{\beta} \cdot \vec{x} \right)}~,&
\Lambda &= 2\pi (M^{-1} \vec{\beta} ) \cdot d\vec{x}~,
\end{align}
and find
\begin{align}
\cA^1 & = 2\pi \left(\ff{\beta}{N} + \vec{\beta}\cdot \vec{x}\right)  \left( dx_{d+1} - (M^{-1} \vec{k}) \cdot d\vec{x}\right)~.
\end{align}
Because $\Theta$ is not an integrally quantized form, the requirement that $\ff{1}{2\pi} d\cA^1$ is an integral form on $\Ttor^{d}$ is non-trivial.  Up to gauge transformations of the bundle $L$, we find that the only solutions are: $\vec{\beta} = 0$ or $\vec{k} = 0$  (any $N$), or $N=2$ and $\vec{\beta} = \vec{k}$.  For any such solutions we have
\begin{align}
\ff{1}{2\pi} d\cA^1 & = \vec{\beta} \cdot d\vec{x} \wedge  dx_{d+1}~.
\end{align}

To explore the remaining consistency conditions, we will denote by $\cA^p$  the connection corresponding to the group element $a^p$, with $0 \le p < N$, and similarly for the diffeomorphism $\varphi_p$.  Writing a sum of two such integers as $p_1+p_2 = m_1 + m_2 N$, with $0 \le m_1 <N$, consistency with the group law~(\ref{eq:babyBA}) requires
\begin{align}
\cA^{m_1} - \cA^{p_1} - \varphi^\ast_{p_1} ( \cA^{p_2}) = i d\log h^{p_1,p_2}~.
\end{align}
Using the explicit form of the $\cA^p$ we find 
\begin{align}
\label{eq:Znbundles}
\cA^{m_1} - \cA^{p_1} - \varphi^\ast_{p_1} ( \cA^{p_2})  = -2\pi m_2 \beta \Theta = i m_2 d\log f~,
\end{align}
where
\begin{align}
f =  \exp  \left[ -2\pi i \beta \left(x_{d+1} - (M^{-1} \vec{k}) \cdot \vec{x} \right) \right]~.
\end{align}
%
%
To describe the $h^{p_1,p_2}$, we define
\begin{align}
\sigma(p) &= \begin{cases}  0~, & p < N~;\\ 1~, &p \ge N~. \end{cases}~, &
\htld(p) &= f^{\sigma(p)}~,
\end{align}
as well as a remainder $\pbar$ such that $\pbar = p \mod N$, and $0 \le \pbar < N$.  With that preparation, we can now write the general solution for the $h^{p_1,p_2}$ as
\begin{align}
h^{p_1,p_2} = \htld(p_1+p_2) c^{p_1,p_2}~,
\end{align}
where the $c^{p_1,p_2}$ are constants.  With this form the constraint~(\ref{eq:hconsistency}) is
\begin{align}
\frac{\varphi^\ast_{p_3} ( \htld(p_1+p_2)) \htld(\overline{p_1+p_2}+p_3)}
{\htld(p_1 + \overline{p_2+p_3}) \htld(p_2+p_3)} = 
\frac{c^{p_1,\overline{p_2+p_3}} c^{p_2,p_3}}{c^{p_1,p_2} c^{\overline{p_1+p_2},p_3}}~.
\end{align}
Using the explicit form of $f$, we find
\begin{align}
\varphi_{p_3}^\ast( \htld(p) ) = \exp\left[-2\pi i \ff{\beta k}{N}p_3 \sigma(p) \right] \htld(p)~,
\end{align}
so that~(\ref{eq:hconsistency}) requires
\begin{align}
\frac{\htld(p_1+p_2) \htld(\overline{p_1+p_2}+p_3)}
{\htld(p_1 + \overline{p_2+p_3}) \htld(p_2+p_3)} = \exp\left[2\pi i \ff{\beta k}{N} \sigma(p_1+p_2)p_3 \right]\frac{c^{p_1,\overline{p_2+p_3}} c^{p_2,p_3}}{c^{p_1,p_2} c^{\overline{p_1+p_2},p_3}}~.
\end{align}
The reader can check that the left-hand-side is not only $x$-independent, but in fact equals $1$.  It is also possible to show that the phase $\exp\left[2\pi i \ff{\beta k}{N} \sigma(p_1+p_2)p_3 \right]$ gives a non-trivial cocycle in $H^3(\Z_N,\GU(1)) \simeq \Z_N$ whenever $\beta k \neq 0 \mod N$, in which case it is not possible to choose the $c^{p_1,p_2}$ to satisfy the consistency conditions.\footnote{This is easy to see explicitly for $N=2$: setting $p_1= p_3=1$ and $p_2 =0$, we find the requirement $c^{0,1} = c^{1,0}$, while taking $p_1 = p_2 = p_3 =1$ leads to $e^{i \pi \beta k} c^{1,0}=c^{0,1}$.
}
 On the other hand, when $\beta k = 0 \mod N$, $c^{p_1,p_2}$ must be a cocycle, and since $H^2(\Z_N, \GU(1)) = 1$, it must in fact be a group coboundary, so that we can set $c^{p_1,p_2} = 1$ by constant gauge transformations on the individual bundles $\cL^a$.

Finally, we should take care of an ambiguity in our description: the parameters $(k,\vec{k})$ and $(\beta,\vec{\beta})$ are not uniquely defined.  For any $\vec{r} \in \Z^4$ there are changes of coordinates on $T^4$
\begin{align}
\vec{x} \to \vec{y} &=  \vec{x} + \Mt^{-1} \vec{r}~, &~ \Mt & = \varphi_a -\iden~,
\end{align}
which preserve the $\Z_N$ action on $\Ttor^4$, i.e. $a(\vec{y}) = \varphi_a(\vec{y}) + \vec{r} \sim \varphi_a(\vec{y})$ and lead to shifts in the parameters $\beta$ and $k$:
\begin{align}
k & \to k + N \vec{k}\cdot \Mt^{-1} \vec{r} ~, &
\beta & \to \beta + N \vec{\beta} \cdot \Mt^{-1} \vec{r}~,
\end{align}
and these shifts do not in general preserve $\beta k \mod N$.  Since the consistency of the equivariant gerbe should not depend on which representative we choose, we demand that the condition $\beta k = 0 \mod N$ is preserved for all $\vec{r}\in \Z^4$.

\subsubsection*{Summary: vertical flat $\Z_N$--equivariant gerbes on $\Ttor^5$}
With these results in hand, we can now use the language of correspondence spaces to describe the possible $\Z_N$--equivariant flat vertical gerbes on $\Ttor^{5}$ with $\Z_N$ action generated by~(\ref{eq:ourZN}).  Each correspondence space $Z_{k,\vec{k};\beta,\vec{\beta}}$ is a double circle fibration over $\Ttor^{d}/\Z_N$ with action generated by
\begin{align}
a(\vec{x},x_5,x_6) = (\varphi_a(\vec{x}), x_5 + \ff{k}{N} + \vec{k} \cdot \vec{x}, x_6 + \ff{\beta}{N} + \vec{\beta} \cdot \vec{x})~.
\end{align}
subject to conditions
\begin{enumerate}
\item $( \beta + N \vec{\beta} \cdot \Mt^{-1} \vec{r}) ( k+ N \vec{k}\cdot \Mt^{-1} \vec{r})  = 0 \mod N$ for all $\vec{r} \in \Z^4$~;
\item $\vec{k} =0$ or $\vec{\beta} = 0$ for $N\ge2$, or $\vec{\beta} = \vec{k} \neq 0$ for $N=2$.
\end{enumerate}

There is a geometric perspective on these consistency conditions.  The $\Z_N$--equivariant gerbe on $\Ttor^5$ defines a gerbe on the quotient $X = \Ttor^5/\Z_N$, and for flat gerbes this amounts to an assignment of holonomies to homology $2$-cycles in $X$, subject to the requirement that any two homologous cycles must be assigned the same holonomy.  Using the computations of holonomies assigned to cycles in $\Ttor^4/\Z_N$ by equivariant line bundles given in~\cite{deBoer:2001px}, we checked that 
our two consistency conditions are the ``upstairs'' view of this requirement.   

We can also understand the conditions from the point of view of the orbifold CFT.  As we will see shortly, and as is discussed in greater length in~\cite{Sharpe:2003cs} and reviewed in detail in the appendix, a choice of gerbe amounts to dressing the orbifold twisted sectors with additional phases, which are constrained by unitarity, modular invariance, and factorization properties of the CFT~\cite{Vafa:1986wx}.

\subsection{T-duality}
Having described a class of $\Z_N$--equivariant flat gerbes on $\Ttor^5$, we return to our original question:  what is the T-dual of type II compactification on $X_{k,\vec{k}}$?  The answer is captured by the correspondence space.  In the previous section we constructed what one might call ``T-dualizable'' flat gerbes on $\cG_{\beta,\vec{\beta}} \to X_{k,\vec{k}}$, and the geometries fit into the correspondence space
\begin{equation}
\begin{tikzcd}
~ & Z_{k,\vec{k};\beta,\vec{\beta}} \ar[ld, "p_g"] \ar[rd,"p_b"'] &~  \\
X_{k,\vec{k}}  \ar[rd, "\pi_g"] \ar[rr,"\text{T-dual}",leftrightarrow]&~ & X_{\beta,\vec{\beta}} \ar[ld,"\pi_b"'] \\
~  & \Ttor^4/\Z_N& ~
\end{tikzcd}
\end{equation}
As we reviewed above, a key result in the study of T-duality with $H$--flux is that $H$ is T-dualizable if and only if the dual flux $\Hh$ is T-dualizable~\cite{Bouwknegt:2003vb}.  The consistency conditions on the gerbes are also symmetric under the exchange $(k,\vec{k}) \leftrightarrow (\beta,\vec{\beta})$.

So, we find that the smooth geometry $X_5 = X_{1,0}$ with trivial gerbe is T-dual to the singular geometry $\Ttor^4/\Z_N\times S^1$ equipped with a flat vertical gerbe with a $\Z_N$--equivariant description on $\Ttor^5$ with $\beta = 1$ and $\vec{\beta} = 0$.  

Similarly, if we start with $X_{0,0} = \Ttor^4/\Z_N\times S^1$ but now turn on a gerbe with $\beta = 0$ and $\vec{\beta}\neq 0$, the T-dual geometry is $X_{0,\vec{\beta}}$ with trivial gerbe.  While both spaces are singular, the singularity structure is quite different.  For example, in the $N=2$ case the singular locus of $X_{0,0}$ consists of $(\vec{x},x_5)$, where $x_{1,2,3,4} \in \{0,\ff{1}{2}\}$, while $x_5$ is arbitrary.  On the other hand, the singular locus for $X_{0,\vec{\beta}}$ requires that in addition $x_4=0$:  instead of $16$ singular circles, the are just $8$.

\subsection{Equivariant gerbes and orbifold CFT}
So far we have defined a class of gerbes and used the correspondence space to suggest a class of T-dual orbifold geometries.  In this section we will connect that perspective with an explicit worldsheet computation.  

The notion that equivariant gerbes enter the construction of the orbifold partition function was developed in~\cite{Sharpe:2000ki,Sharpe:2003cs},  and the second reference showed that pure holonomy equivariant gerbes with $B = 0$ produce the kinds of shift orbifold factors that are necessary to relate the $X_5$ and $\Ttor^4/\Z_N\times S^1$ orbifolds.   We will now review those ideas and illustrate in an example that they do lead to the expected phases.

Let $Z_{0,0}$ denote the genus $1$ partition function of a CFT which admits the action of a discrete abelian group $G$ with $0$ the additive identity element.  The orbifold partition function then takes the familiar form
\begin{align}
\cZ &= \frac{1}{|G|}\sum_{a,b\in G}  Z_{a,b}~, &  Z_{a,b} &= \Tr_{\cH_b} 
\rho_b(a) \bq^{L_0 -c/24} \bqbar^{\Lb_0-\cb/24} ~,
\end{align}
where $\rho_b(a)$ is a representation of the $G$-action on the Hilbert space $\cH_b$, i.e. the space of states twisted by the element $b$.  In general the choice of representation $\rho_b$ is not unique.  For example, if $\rho_b(a)$ leads to a consistent choice of modular-invariant and unitary partition function, then so does  
\begin{align}
\rhot_b(a) = \rho_b(a) \frac{\omega^{a,b}}{\omega^{b,a}}
\end{align}
for any representative of a class $[\omega] \in H^2(G,\GU(1))$---this is the discrete torsion of~\cite{Vafa:1986wx}. 

The observation from~\cite{Sharpe:2000ki} is that the equivariant gerbe data  for a $G = \Z_N$ quotient enters the choice of $\rho_b(a)$ through the pull-back of the holonomy of the gerbe to the string worldsheet:
\begin{align}
\label{eq:EricsPhase}
P_{a,b} = \exp\left[ i \int \Phi^\ast(B) \right] \exp\left[ i\int_{0}^{1} \Phi^\ast(\cA^b)- i\int_{0}^{\tau} \Phi^\ast(\cA^a)  \right] \frac{h^{a,b}(x)}{h^{b,a}(x)}~
\end{align}
in the Lagrangian computation of $Z_{a,b}$.  Here $\Phi : T^2 \to X$ is the map from the worldsheet torus with complex structure $\tau =\tau_1+i\tau_2$ to the targetspace with
\begin{align}
\Phi(0) &= x~, &
\Phi(1) & = \varphi_b(x)~,&
\Phi(\tau) & = \varphi_a(x)~,&
\Phi(1+\tau) & = \varphi_{ab} (x)~.
\end{align}
The integration contours in the $z$ plane of the worldsheet torus are along the left and bottom boundaries of the parallelogram:
\begin{equation}
\begin{tikzpicture}
\draw[thin] (0,0) node [below] {$0$} -- (2,0) node [below] {$1$}  -- (3,1) node[right] {$\tau+1$} -- (1,1) node[above] {$\tau$} --cycle;
\draw[thin] (5,1.3) --++(0,-0.2)--++(0.2,0) node[above] {$z$};
\end{tikzpicture}
\end{equation}
The expression $P_{a,b}$ is not easy to interpret on a general targetspace.  However, our interest is in $X = \Ttor^d/\Z_N \times S^1$, where the interpretation is straightforward in the situation when $B = 0$.  Focusing on the bosonic part of the CFT, the original theory's configuration space decomposes into sectors labeled by their periodicities under shifts of $z\to z+1$ and $z\to z+\tau$.  The $x_5$ field corresponding to $S^1$, which is neutral under the orbifold action, has configurations
\begin{align}
\label{eq:semi-class}
x_5 (z,\zb) & = \ff{i}{2\tau_2} \left( w- \nt\taub\right) z 
+\ff{i}{2\tau_2} \left(-w+\nt\tau\right)\zb~.
\end{align}
While $w$ has the interpretation as the winding quantum number, $\nt$ is not the momentum mode $n$ of the Hamiltonian formulation---the two are related through a Poisson resummation.  In constructing the path integral of $Z_{a,b}$ we must also consider fluctuations around this classical solution, but those do not enter the phase factor $P_{a,b}$.  

In our examples above, the gerbes with $B = 0$ have $\cA^a = \ff{2\pi a}{N} dx^5$ and $h^{a,b} = h^{b,a}$, which leads to phases
\begin{align}
P_{a,b} = \exp\left[\ff{2\pi i}{N} \left( aw -b\nt\right)\right]~.
\end{align}
The contribution to the circle partition function can be evaluated by including the integration over the fluctuations around the semi-classical solution~\cite{Ginsparg:1988ui}.  Including the phase $P_{a,b}$ the result is (our conventions are given in appendix~\ref{app:bosonicorbifold})
\begin{align}
Z^{\text{cir}}_{a,b} = \ff{1}{\eta\etab} \sum_{w} e^{2\pi i \ff{a w}{N}} \ff{\sqrt{\tau_2}}{r} \sum_{\nt} \exp\left[ -S_0(\nt,w,\tau,r) - 2\pi i \ff{b}{N}\nt\right]~,
\end{align}
where $S_0$ is the classical action evaluated on the solution~(\ref{eq:semi-class}).
Making a Poisson resummation on $\nt$, we therefore obtain
\begin{align}
Z^{\text{cir}}_{a,b} = \ff{1}{\eta\etab} \sum_{w} e^{2\pi i \ff{a w}{N}} \ff{\sqrt{\tau_2}}{r} \sum_{n}
\int_{-\infty}^{\infty} dx \exp\left[ -S_0(x,w,\tau,r) - 2\pi i \ff{b}{N}x -2\pi i x n\right]~.
\end{align}
When $b =0$, this leads to the usual expression for the circle partition function with an extra phase:
\begin{align}
Z^{\text{cir}}_{a,0} = \ff{1}{\eta\etab} \sum_{n,w} e^{2\pi i \ff{a w}{N}}  \bq^{h_{n,w}} \bqbar^{\hb_{n,w}}~,
\end{align}
but now the form of the $b$-dependence under the $x$ integral shows that more generally including $b$ just leads to the shift $n\to n + b/N$ in the weights $h_{n,w}$ and $\hb_{n,w}$.  But this is precisely the partition function of the T-dual of the shift orbifold of the circle!  Since the remaining degrees of freedom, i.e. the fermions and the bosons of the $\Ttor^d$ theory, are treated in the same way in both the $X_5$ and $\Ttor^d /\Z_N \times S^1$ orbifolds, the gerbe data and the associated phase correctly reproduce the T-dual of the $\Z_N$ quotient that leads to the smooth $X_5$ geometries.

It would be interesting to generalize this treatment to the other pure holonomy equivariant gerbes that appear in the T-dualities, i.e. the ones with $B\neq 0$.  On one hand, it would provide a detailed test of the dualities constructed via the correspondence space analysis.  On the other hand, such a generalization may shed light on the precise meaning of~(\ref{eq:EricsPhase}) in compactifications that go beyond the simple toroidal examples considered here.  We leave such a study for future work.

\subsubsection*{The circle and ``non-geometry''}
Much like the inexhaustible electron~\cite{Bunge:1950ie}, the compact boson is a seemingly perpetual source of insights and lessons, and we end this section by specializing our analysis to it.

The compact boson CFT admits various orbifold actions, and some of these, such as the shift orbifold which acts by the phase $e^{2\pi i n /N}$ or the reflection orbifold, have a straightforward geometric interpretation:  for example, the former is a quotient that leads to a circle of radius $r/N$.  On the other hand, there are also orbifolds that seemingly do not have such a straightforward interpretation, such as the $\Z_N$ orbifold that acts by the phase $e^{2\pi i w/N}$. If we are willing to include the gerbe data as part of our geometric information, then we can assign a geometric meaning to this phase:  it corresponds to turning on a gerbe on $S^1$!

Applying this logic to the compact boson CFT is surely an indication that one has spent too much time with gerbes, but the example has an important lesson for more elaborate theories:  the gerbe data should be included in any discussion of ``non-geometric'' features of string compactification.  

\subsection{Frozen singularities and a dual perspective}
The geometry $\Ttor^4/G \times S^1$ is familiar to every string theorist as a limit of compactification on $\text{K3} \times S^1$.   The resulting spacetime physics depends on the way in which the limit is taken.  For example, if we focus on type II compactification, then the choice of $B$-field determines whether the limiting theory will lead to additional massless gauge degrees of freedom (in IIA) or tensionless strings (in IIB)~\cite{Witten:1995zh,Aspinwall:1995zi,MR1416354}.  These arise at points in the moduli space where the CFT breaks down, and string non-perturbative effects are crucial.  On the other hand, there is a choice of $B$-field such that the limit leads precisely to the orbifold CFT for $\Ttor^4/G$.  In each of these cases it is possible to perturb the moduli to a more generic point and arrive at a smooth K3 geometry with some small cycles.  From the point of view of the orbifold CFT these deformations correspond to massless spacetime fields that arise from the twisted sectors.

As we have seen, in $5$--dimensional compactification there is a new possibility:  by turning on a flat vertical gerbe on the $\text{K3} \times S^1$ geometry, we obtain a theory without moduli associated to any of the blow-up modes.   The disappearance of the moduli is easy to understand in the T-dual picture, since the dual geometry is smooth, and every twisted sector is massive.   In the original formulation the singularity is frozen by the holonomy of the gerbe.  

We can understand this as follows in terms of the local structure of the geometry.\footnote{In this discussion we stick to the equivariant gerbe with $B=0$.}  The equivariant gerbe implies that the orbifold geometry $\Ttor^4/G \times S^1$ carries non-trivial gerbe holonomy: a curve $\gammat$ in $\Ttor^4$ with $\gammat(0) =x$ and $\gammat(1) = \varphi_a(x)$ projects to a closed curve $\gamma$, and the cycle $\gamma \times S^1$ then carries a gerbe holonomy $e^{2\pi i a/N}$:  this topological feature is responsible for freezing the moduli.  

This local discussion has a string-dual description:  starting with IIB compactified on $\Ttor^4/G\times S^1$ with a $B=0$ vertical gerbe, we use $\SL(2,\Z)$ duality to obtain IIB compactified on the same space, but now with a holonomy for the Ramond-Ramond $C_2$ field on the same cycle $\gamma \times S^1$.  Taking another T-dual, we obtain IIA on $\Ttor^4/G \times S^1$, but now with holonomy for $C_1$ on the loop $\gamma$.  These different descriptions can understood by compactifying M-theory on $X_5 \times S^1$ and then reducing either on the trivial circle or on the circle of the $X_5$ fibration.  This is exactly the picture described in~\cite{deBoer:2001wca} for the freezing of the singularities, and we see that our construction gives the same mechanism at the level of spacetime physics, but one accessible in standard worldsheet analysis.

If we consider the $\Ttor^4/\Z_2 \times S^1$ case,  there are two qualitatively different gerbes we might turn on:  the $B=0$ one is dual to the smooth geometry $X_5 = X_{1,0}$, while the $B\neq0$ one is T-dual to compactification on the singular space $X_{0,\vec{k}}$.  In the former the gerbe freezes all $16$ singularities, while in the latter only $8$ are frozen, and the remaining $8$ contribute massless moduli that can be used to resolve the singularities.\footnote{Note that such a resolution in the $X^2_{0,\vec{k}}$ geometry provides examples of the recent construction of novel heterotic geometries in~\cite{Fino:2019mvp}.}  

This leads to a natural question:  is it possible to turn on different gerbe holonomies at different singularities to freeze a number of singularities different from $0$, $8$, or $16$?  As pointed out in~\cite{deBoer:2001wca}, there are global constraints that prevent us from turning on such holonomies independently:  some combinations of the cycles are homologous, and unless the holonomies are chosen in a consistent manner, they would necessarily lead to non-zero curvature for $dC_1$.

These global constraints are also reflected in our worldsheet analysis.  This can be seen explicitly by focusing on the bosonic sector of the theory.  For example, the partition function for $\Ttor^{4}/\Z_2 \times S^1$---see appendix~\ref{app:bosonicorbifold} for the relevant CFT details---has twisted sector contributions that come with an overall factor of $16=2^{4}$, one for each of the orbifold fixed points:
\begin{align}
\cZ_{\text{twist}} = 2^{4} Z^{\text{cir}}(r) \left\{ \left({\frac{\eta\etab}{\vartheta_4 \bar{\vartheta}_4}}\right)^{2} + \left({\frac{\eta\etab}{\vartheta_3 \bar{\vartheta}_3}}\right)^{2} \right\}~.
\end{align}
A CFT model of turning on a local gerbe on $k$ of the singularities is to split the twisted sector contribution into $k$ terms for which we introduce the phases associated to the gerbe, and the remaining $16-k$ terms which we leave untouched.  Thus, we have
\begin{align}
\cZ^{(k)}_{\text{twist}}  &= (16-k) Z^{\text{cir}}(r) \left\{ \left({\frac{\eta\etab}{\vartheta_4 \bar{\vartheta}_4}}\right)^{2} + \left({\frac{\eta\etab}{\vartheta_3 \bar{\vartheta}_3}}\right)^{2} \right\}\nonumber\\
&\quad + k \left\{Z_{\text{sh}-}^{~\,\,+} \left({\frac{\eta\etab}{\vartheta_4 \bar{\vartheta}_4}}\right)^{2} + Z_{\text{sh}-}^{~\,\,-}\left({\frac{\eta\etab}{\vartheta_3 \bar{\vartheta}_3}}\right)^{2} \right\}~.
\end{align}
Can this be the twisted sector of a worldsheet $\Z_2$ orbifold CFT?  If so, we should be able to obtain the untwisted sector partition function by applying worldsheet $\SL(2,\Z)$ transformations.  However, when we do this, we find that the resulting expression fails to have an integral $\bq$ expansion unless $k = 0 \mod 8$.

A similar analysis can be carried out for the $\Z_3$ example: there the number of frozen singularities can be $0,6,9$~\cite{deBoer:2001wca}, and those are precisely the values for which the procedure just outlined yields a well-behaved worldsheet partition function.

\section{Discussion}
It is remarkable that after so many years of intensive study theories with $8$ supercharges continue to furnish surprises and provide insights into the structure of quantum field theory and string theory.  In this work we have seen new connections between disparate geometries, both in the context of heterotic and type II theories in $5$ dimensions.  In the former we found that duality in the $\GO(21,5)$ moduli space leads to relations between topologically distinct compactifications.  In the latter, we saw that by including the global data of a pure holonomy gerbe, T-duality is enlarged beyond the conventional arena of Buscher rules and exchange of data encoded in the differential forms associated to a circle fibration equipped with $H$-flux.

The original motivation for our work was to understand how a particularly simple class of compact flat geometries fits into the framework of string compactification.   We showed that such compactifications are related to more familiar ones, and, via T-duality, they provide models for a number of phenomena.  Perhaps the most striking of these is the freezing of singularities in a $\Ttor^4/\Z_N\times S^1$ compactification achieved by turning on a pure holonomy gerbe.  Such phenomena arise in a number of settings, often involving strongly coupled physics or other degrees of freedom that are difficult to access directly (e.g. background Ramond-Ramond fields).  Our work provides an example where the mechanism can be understood directly in the worldsheet theory and has a geometric interpretation via the gerbe.

There are a number of promising directions where extensions of our ideas may be of use, and in the remainder of this section we will highlight a few of them.

First, we have seen that inclusion of the gerbe data leads to new classes of and new equivalences between orbifold conformal field theories.  While our analysis was carried out in the context of $8$ supercharges and $5$ dimensions, it should be possible to extend our techniques to other orbifolds, and the classification of~\cite{Fischer:2012qj} offers an obvious place to start.\footnote{It appears that the  type II $6$-dimensional orbifolds that preserve $16$ supercharges do not allow non-trivial equivariant gerbes beyond those explored in this work.  However, we have not explored the situation with type II orbifolds that only preserve $8$ supercharges.}

  On one hand, inclusion of the pure holonomy gerbes will enlarge the class of conformal field theories, but, on the other hand, there will also be new equivalences between geometrically distinct orbifolds equipped with the gerbe data.  For general orbifold groups  the possibility of discrete torsion should also be included in a general analysis.

Sticking to the $5$--dimensional theories, perhaps the simplest extension is to consider replacing the base geometry in our circle fibrations to $\text{K3}/G$.  Although it will not be possible to solve the worldsheet theory exactly, it is possible to preserve $8$ supercharges, and therefore to have a reasonable amount of control on the theory.  At the same time, the non-trivial topology of K3 opens up new possibilities.  For example, in the context of the heterotic string there is the possibility of choosing a principal circle fibration that has a non-torsion first Chern class~\cite{Dasgupta:1999ss,Fu:2006vj,Becker:2006et,Melnikov:2012cv}.  Generalizing such fibrations to include a shift in the $S^1$ leads to a large class of examples where one can explore the topology and geometry of gerbes and perhaps find new dualities relating these background to each other beyond those obtained by conventional T-duality with $H$-flux~\cite{Bouwknegt:2003vb,Evslin:2008zm}.  

A more ambitious direction would be to pursue our ideas in the context of the recent constructions~\cite{Fino:2019mvp} that generalize the geometries of~\cite{Fu:2006vj} by allowing the base K3 geometries to have a number of orbifold singularities:  the total space of the fibration is nevertheless smooth, precisely because the local geometry near each singularity is $(\C^2 \times S^1)/G$, where the $G$-action on $\C$ is accompanied by a shift on the $S^1$ fiber.  Thus, these are models without a global $(\text{K3}\times S^1)/G$ orbifold description.  It will be interesting to explore the interplay of the differential geometry and gerbe holonomy in this context.  Is it possible to obtain ``local'' T-dual descriptions relating the smooth $(\C^2 \times S^1)/G$ geometry to a singular $\C^2/G \times S^1$ with a  non-trivial gerbe holonomy?  What are the global constraints on the set of singularities that can be treated in this fashion?

Returning to the simpler $\Ttor^4$ geometry as the starting point, we can study how our results fit into type II/ heterotic duality.  Since the $6$-dimensional duality between type IIA on K3 and heterotic string on $\Ttor^4$ is well-understood, we can hope to use the spacetime perspective, where we view heterotic compactification on $X_5$ as a reduction of the $6$-dimensional theory on a circle with a non-trivial holonomy for the $G$ action, to map our results to type II.  Can we interpret the dual type II theory as an orbifold?  How do such dual pairs relate to the work~\cite{Hull:2017llx,Gautier:2019qiq} on non-geometric type II compactifications and their heterotic duals?  More generally, what is the translation of the gerbe data in either duality frame to the other one?

Finally, our work shows that inclusion of the global gerbe data is crucial in understanding T-duality, and even our simple examples indicate that much remains to be understood.  We indicated some of the directions in the main text, as well as in appendix~\ref{app:equivariantgerbes}.  Another direction is to consider our results in the context of heterotic T-duality, where the gerbe data is more complicated, and there can be interesting interplay between discrete torsion and more conventional shift orbifolds~\cite{Ploger:2007iq}.  It would be very interesting to understand these phenomena in the context of a heterotic correspondence space~\cite{Evslin:2008zm}.

\appendix

\section{Bosonic toy model} \label{app:bosonicorbifold}
In this appendix we discuss the bosonic partition function for compactification on $X_5$ in the case of the $\Z_2$ quotient.  This will illustrate some features of the decompactification limit to six dimensions discussed in section~\ref{ss:X5compactification}, and it will also set up conventions for some later developments. While the $\Z_2$ quotient is particularly simple, there is a straightforward to the remaining $\Z_N$ actions.

\subsection{Orbifolds of $S^1$} 
We described the construction of the shift orbifold of the circle in section~\ref{ss:X5compactification}.   Starting with the circle partition function
\begin{align}
Z(r) &= \frac{1}{\eta\etab} \sum_{n,w\in \Z} \bq^{h_{n,w}} \bqbar^{\hb_{n,w}}~, &
\eta & = \bq^{1/24} \prod_{n=1}^{\infty} (1-\bq^n)~,
\end{align}
where
\begin{align}
\hb_{n,w} = h_R(p) &= \left( \ff{n}{2r} + \ff{rw}{2}\right)^2~,&
h_{n,w} = h_L(p) & =\left( \ff{n}{2r} - \ff{rw}{2} \right)^2~,
\end{align}
the partition function of the shift orbifold for the $\Z_2$ quotient is
\begin{align}
Z_{\text{sh}}(r) = \ff{1}{2} \left ( Z_{\text{sh}+}^{~\,\,+} + Z_{\text{sh}+}^{~\,\,-} +  Z_{\text{sh}-}^{~\,\,+} + Z_{\text{sh}-}^{~\,\,-} \right)~, 
\end{align}
with
\begin{align}
Z_{\text{sh}+}^{~\,\,+} & = Z(r)~, &
Z_{\text{sh}+}^{~\,\,-}  & = \frac{1}{\eta\etab} \sum_{n,w\in \Z} e^{i\pi n} \bq^{h_{n,w}}\bqbar^{\hb_{n,w}}~, \nonumber\\
Z_{\text{sh}-}^{~\,\,+}  & = \frac{1}{\eta\etab} \sum_{n,w\in \Z} \bq^{h^-_{n,w}}\bqbar^{\hb^-_{n,w}}~&
Z_{\text{sh}-}^{~\,\,-}  & = \frac{1}{\eta\etab} \sum_{n,w\in \Z} e^{i\pi n} \bq^{h^-_{n,w}}\bqbar^{\hb^-_{n,w}}~,
\end{align}
where the weights in the twisted sector are
\begin{align}
\label{eq:twistedshiftedweights}
\hb^-_{n,w}  &= \left( \ff{n}{2r} + \ff{r}{2} (w+\ff{1}{2})\right)^2~,&
h^-_{n,w} & =\left( \ff{n}{2r} - \ff{r}{2} (w+\ff{1}{2}) \right)^2~.
\end{align}
This is just a fancy rewriting of $Z(r/2)$, but the split into the sectors will be useful for building the $X_5$ orbifold CFT.

\subsubsection*{A reflection orbifold of $S^1$}
Another familiar quotient of the compact boson is the $x\to -x$ orbifold.  In this case the action does affect the oscillators, and it acts on the states $|p\ra$ by $g' |p \ra = |-p\ra$.\footnote{Note that our discussion of cocycles and phases in section~\ref{ss:cocycles} shows that there is no additional phase on the state.}  As discussed in~\cite{Ginsparg:1988ui}, the orbifold partition function can be obtained by explicitly constructing the untwisted sector with the $g'$ projection, and the the twisted sector contributions are obtained by taking orbits of $\SL(2,\Z)$.  The result is
\begin{align}
Z_{\text{re}}(r) =  \ff{1}{2} \left ( Z_{\text{re}+}^{~\,\,+} + Z_{\text{re}+}^{~\,\,-} +  Z_{\text{re}-}^{~\,\,+} + Z_{\text{re}-}^{~\,\,-} \right)~,
\end{align}
with
\begin{align}
Z_{\text{re}+}^{~\,\,+} & = Z(r)~, & 
Z_{\text{re}+}^{~\,\,-} &= 2 \sqrt{\frac{\eta\etab}{\vartheta_2 \bar{\vartheta}_2}}~,&
Z_{\text{re}-}^{~\,\,+} &= 2 \sqrt{\frac{\eta\etab}{\vartheta_4 \bar{\vartheta}_4}}~,&
Z_{\text{re}-}^{~\,\,-} &= 2 \sqrt{\frac{\eta\etab}{\vartheta_3 \bar{\vartheta}_3}}~.
\end{align}
Our conventions for the Jacobi theta functions $\vartheta_i$ are as in~\cite{Ginsparg:1988ui}.  The crucial factors of $2$ in the twisted sector contributions reflect the presence to two fixed points for the orbifold action, at $x = 0$ and $x = 1/2$.

\subsection{Bosonic CFT for $X_5$} 
To make contact with our compactification of $X_5$, we first generalize the $S^1$ reflection orbifold to a reflection orbifold of $\Ttor^d$.  Since the reflection symmetry is present for all parameters of the $\Ttor^d$ CFT, we have
\begin{align}
Z_{\text{re}+}^{~\,\,+} & = Z_{\Ttor^d}~, & 
Z_{\text{re}+}^{~\,\,-} &= 2^d \left({\frac{\eta\etab}{\vartheta_2 \bar{\vartheta}_2}}\right)^{d/2}~,&
Z_{\text{re}-}^{~\,\,+} &= 2^d \left({\frac{\eta\etab}{\vartheta_4 \bar{\vartheta}_4}}\right)^{d/2}~,&
Z_{\text{re}-}^{~\,\,-} &= 2^d \left({\frac{\eta\etab}{\vartheta_3 \bar{\vartheta}_3}}\right)^{d/2}~.
\end{align}
Finally, to obtain the bosonic CFT for $X_5$, we combine the orbifold actions on $\Ttor^{d} \times S^1$ and find the orbifold partition function
\begin{align}
\cZ & =  \ff{1}{2} \left(\cZ_+^+ + \cZ_+^- + \cZ_-^+ + \cZ_-^-\right)~,
\end{align}
with
\begin{align}
\cZ_+^+ & = Z(r) Z_{\Ttor^d}~,&
\cZ_+^-  & = Z_{\text{sh}+}^{~\,\,-}Z_{\text{re}+}^{~\,\,-}~,&
\cZ_-^+  & = Z_{\text{sh}-}^{~\,\,+}Z_{\text{re}-}^{~\,\,+}~,&
\cZ_-^-  & = Z_{\text{sh}-}^{~\,\,-}Z_{\text{re}-}^{~\,\,-}~.
\end{align}

\subsubsection*{A decompactification limit}
Having written down the full bosonic partition function for $X_5$ (or rather for its $d+1$-dimensional analog), we are now ready to examine its decompactification limit:  $r\to \infty$.  

To get an idea of what we mean, we first consider the circle partition function itself.  The dimensions of the primary states $|p\ra$ are 
\begin{align}
\Delta_{p} = h_L(p) + h_R(p) = \frac{n^2}{2r^2} + \frac{w^2 r^2}{2}~,
\end{align}
so that when $r$ is very large, every state with $w \neq 0$ has a large dimension, and therefore has the spacetime interpretation of a very heavy mode.  To focus on a set of light states, we fix a cut-off $\Delta_\ast$, and study the contribution to the partition function from states with $\Delta \le \Delta_\ast$.   We define Kaluza-Klein (KK) tower sums 
\begin{align}
\boldsymbol{T}^{\pm}_{\Delta_\ast} = \sum_{\Delta(p) \le \Delta_\ast} (\pm 1)^n \bq^{h_{n,w}} \bqbar^{\hb_{n,w}}~.
\end{align}
So, for example, if we set $\Delta_\ast = 2$, we find
\begin{align}
Z(r)  &= (\bq \bqbar)^{-1/24} \left( \sum_{n^2\le 4r^2} (\bq\bqbar)^{n^2/4r^2} +(\bq+\bqbar) \sum_{n^2\le 2 r^2} (\bq\bqbar)^{n^2/4r^2} + 2\bq^2 + \bq \bqbar + 2\bqbar^2
\right) + \text{heavy}~ \nonumber\\
 & = (\bq \bqbar)^{-1/24} \left( \boldsymbol{T}^+_2 + (\bq+\bqbar)\boldsymbol{T}^+_1 + \bq\bqbar + 2(\bq^2+\bqbar)^2 \right) + \text{heavy}~.
\end{align}
We understand these contributions:
\begin{enumerate}
\item the over all factor is simply the zero-point energy $ (\bq \bqbar)^{-c/24}$ with $c=1$;
\item the $\boldsymbol{T}_2^+$ is the KK expansion of the scalar mode of the circle metric;
\item the  $\boldsymbol{T}_1^+$ is the KK expansion of the gauge bosons from the circle reduction of the metric and $B$-field;
\item the $\bq\bqbar$ reflects the presence of the marginal operator that controls the radius of the circle;
\item the $2\bq^2$ and $2\bqbar^2$ terms are associated to the worldsheet operators $\p\phi\p\phi, \p^2\phi$, and their conjugates.
\end{enumerate}

With a little bit of work, we can obtain a similar expansion with $\Delta_\ast = 2$ for the $\cZ$ partition function.  First, we note that the $w\to w+\ff{1}{2}$ shift in $h^-_{n,w}$ relative to $h_{n,w}$ ensures that all twisted sector states are heavy.  Thus, the light states just come from the untwisted sector.  Making an expansion of the partition function then leads to
\begin{align}
\cZ 
& = (\bq\bqbar)^{-c/24} \left( \frac{1}{2} (\boldsymbol{T}^-_2 + \boldsymbol{T}^+_2)  + (\bq+\bqbar)  \frac{1}{2} (\boldsymbol{T}^-_1 + \boldsymbol{T}^+_1)  + d (\bq+\bqbar) \frac{1}{2} (\boldsymbol{T}^+_1 - \boldsymbol{T}^-_1)\right. \nonumber\\
& \qquad \qquad\qquad\quad \Bigl. +  ( d^2 + 1) \bq\bqbar + \left(\ff{d(d+1)}{2}+2\right) (\bq^2 + \bqbar^2) \Bigr)~ \nonumber\\
&\quad + (\bq\bqbar)^{-c/24} {\sum_{\vec{p}\neq 0} }\bq^{h(\vec{p})} \bqbar^{\hb(\vec{p})} \left[ \frac{1}{2} \boldsymbol{T}^+_{2-\Delta_{\vec{p}}} + (1+d)(\bq+\bqbar) \frac{1}{2} \boldsymbol{T}^+_{1-\Delta_{\vec{p}}} \right] + \text{heavy}~.
\end{align}
Here $c = d+1$.  As in the decompactification of the circle, we can give each of these terms Kaluza-Klein interpretation, but this time there will be a twist in the boundary conditions.
\begin{enumerate}
\item
Scalar KK modes.  These are associated with the factor
\begin{align} \frac{1}{2} (\boldsymbol{T}^-_\Delta + \boldsymbol{T}^+_\Delta)  = \sum_{n \in 2\Z, n^2 \le 4r^2} (\bq\bqbar)^{n^2/4r^2}  = \sum_{n^2 \le 4(r/2)^2 } (\bq\bqb)^{\ff{1}{4}n^2(r/2)^{-2}}
\end{align}
This is interpreted as a KK tower for a periodic field on a circle of radius $r/2$~.
\item Circle Vector KK modes.  These are associated with the factor
\begin{align}
 (\bq+\bqbar)  \frac{1}{2} (\boldsymbol{T}^-_1 + \boldsymbol{T}^+_1)~,
\end{align}
and again have an interpretation of a KK tower for a circle of radius $r/2$, this time dressing the circle's vector field.
\item $\Ttor^d$ Vector KK modes.  These are associated with the factor
\begin{align}
d (\bq+\bqbar) \frac{1}{2} (\boldsymbol{T}^+_1 - \boldsymbol{T}^-_1)
\end{align}
This time we look at a slightly different KK term:
\begin{align}
\frac{1}{2} (\boldsymbol{T}^+_\Delta - \boldsymbol{T}^-_\Delta) =   \sum_{n \in 2\Z+1, \Delta_{m,0} \le 2} (\bq\bqbar)^{n^2/4r^2}  = \sum_{(n+1/2)^2 \le 4(r/2)^2  } (\bq\bqbar)^{\ff{1}{4}(n+1/2)^2 (r/2)^{-2}}~.
\end{align}
Because the sum runs over odd integers, we see that we do not have any massless vectors from the $\Ttor^{d}$ for any finite $r$---this can be contrasted with the circle's vector, which has a massless mode and a KK tower over it.

These states constitute the KK spectrum we would get for a field that is anti-periodic on the circle of radius $r/2$:  i.e. a field satisfying
\begin{align}
A (\theta + 2\pi ) = - A(\theta)~.
\end{align}
Such a field has a Fourier expansion
\begin{align}
A = \sum_n a_n e^{i(n+1/2) \theta}~.
\end{align}

\item The moduli are responsible for the $(d^2 +1)\bq\bqbar$ term, and they give rise to massless spacetime fields.

\item The canonical spin $2$ fields that contribute to $\bq^2$ are also nicely interpreted as  the $d(d+1)/2$ operators $\p x_i \p x_j$ and $\p\phi\p\phi$ and $\p^2\phi$.  In other words, these are the spin $2$ operators invariant under the $\Z_2$ reflection on $\Ttor^{d}$.  The same holds for the  $\bqbar^2$ contributions.

\item The towers over $\Ttor^d$ momentum/winding modes.  These arise from the second to last term that we kept in the expansion, which we rewrite as
\begin{align}
\sum_{\vec{p}\neq 0}\bq^{h(\vec{p})} \bqbar^{\hb(\vec{p})}  \frac{1}{2} \boldsymbol{T}^+_{2-\Delta_{\vec{p}}} = 
\frac{1}{2} \sum_{\vec{p}\neq 0}\bq^{h(\vec{p})} \bqbar^{\hb(\vec{p})}  \left( 
 \frac{1}{2} (\boldsymbol{T}^+_{2-\Delta_{\vec{p}}} +\boldsymbol{T}^+_{2-\Delta_{\vec{p}}})
 +
 \frac{1}{2} (\boldsymbol{T}^+_{2-\Delta_{\vec{p}}} -\boldsymbol{T}^+_{2-\Delta_{\vec{p}}})
 \right)~.
%
\end{align}
These have a clear interpretation.  The $\Ttor^{d}$ states $|\vec{p}\ra$ with $\vec{p} \neq 0$ can be rewritten in terms of
\begin{align}
|\vec{p}\ra^{\pm} = \ff{1}{\sqrt{2}} \left( |\vec{p}\ra \pm |-\vec{p}\ra\right)~.
\end{align}
The $|\vec{p}\ra^+$ states are then dressed with the ``periodic'' KK tower---$\frac{1}{2} (\boldsymbol{T}^+_{2-\Delta_{\vec{p}}} +\boldsymbol{T}^+_{2-\Delta_{\vec{p}}})$, while the $|\vec{p}\ra^-$ states are dressed with the ``anti-periodic'' KK tower---$\frac{1}{2} (\boldsymbol{T}^+_{2-\Delta_{\vec{p}}} -\boldsymbol{T}^+_{2-\Delta_{\vec{p}}})$.

The remaining states have a similar interpretation:  they are just dressed by the oscillator modes coming from $\p x_i$ or $\p\phi_i$ factors and their conjugates.

\end{enumerate}
In summary then, we see in this bosonic model that in the $r\to\infty$ limit, the light states have a clear spacetime interpretation of turning on a holonomy for the $\Z_2$ reflection symmetry acting on $\Ttor^d$:  the invariant fields are expanded in periodic modes on the $S^1$, while the charged fields are expanded in anti-periodic modes.  It should not be too difficult to include fermions in this discussion:  the novel feature that we expect to naturally emerge is the reduction of spacetime supersymmetry due to the non-trivial holonomy.

\section{Equivariant flat gerbes}  \label{app:equivariantgerbes}
In this appendix we review aspects of abelian gerbes and equivariant structures on gerbes, following~\cite{Chatterjee:1998ger,Sharpe:2000ki,Sharpe:2003cs}~.

Consider a compact smooth Riemannian manifold $X$ that admits an action of a finite group $G$: for every $a\in G$ there is a diffeomorphism $\varphi_a : X\to X$, and the composition respects the group structure:  $\varphi_{a} (\varphi_{b} (x)) = \varphi_{ab}(x)$ for all $x \in X$.   It is possible to choose a good cover $\mathfrak{U} = \{U_\alpha\}_{\alpha\in I}$  such that $G$ has an action on the indexing set $I$, with $a : \alpha \to a(\alpha)$, and $\varphi_a(U_\alpha) = U_{a(\alpha)}$  and similarly for all non-empty intersections $U_{\alpha\beta}$, $U_{\alpha\beta\gamma}$, etc.~\cite{YANG2014230}.   We will assume that such a cover has been chosen, so that the various \v{C}ech cochain manipulations that are to follow have a simple interpretation.\footnote{This is not strictly speaking necessary, but it makes for simpler arguments.  A complete treatment would involve also proving that the results are independent of the choice of cover; we will leave that to the references~\cite{Chatterjee:1998ger}.}

\subsection{\v{C}ech cochains and Hermitian line bundles} \label{ss:cechlinebundle}

For what follows, it will be convenient to work in the language of \v{C}ech cochains and the coboundary operator $\delta$, so we will take a moment to review that language and fix our conventions.\footnote{A readable introduction is given in~\cite{Voisin:2007hodge}.}  We consider a sheaf valued in an abelian group $\cF$ defined over our cover $\mathfrak{U}$.  We will denote the cochains by $C^k(X,\cF)$, with $k=0$ denoting the space of sections defined on each $U_\alpha$,  $k=1$ on the $U_{\alpha\beta}$, and so on.  The coboundary operator $\delta$ then takes $\sigma \in C^k(X,\cF)$ to $(\delta\sigma) \in C^{k+1} (X,\cF)$ in the familiar way:  for example, if $\sigma^0 \in C^0(X,\cF)$, and $\sigma^1 \in C^1(X,\cF)$, then we have
\begin{align}
(\delta\sigma^0)_{\alpha\beta}  &=  \sigma^0_\alpha -\sigma^0_\beta~,&
(\delta\sigma^1)_{\alpha\beta\gamma} & = \sigma^1_{\alpha\beta}+\sigma^1_{\beta\gamma} + \sigma^1_{\gamma\alpha}~.
\end{align}
The signs and ordering are chosen so that $\delta^2 =0$.
We say a section $\sigma$ is a cocycle if $\delta\sigma = 0$, and it is a coboundary if it can be written as $\sigma = \delta \lambda$.  Note that we will have occasion to use both additive and multiplicative abelian groups; in the latter case the cocycle condition is written as $\delta \sigma =1$.

The cochains that will show up in our discussion are:
\begin{enumerate}
\item $C^k(X,\Omega^p)$, where $\Omega^p$ denotes smooth $p$-forms;
\item  $C^k(X, S^1)$, where $S^1 = \GU(1)$ denotes circle-valued constants;
\item $C^k(X,\underline{S}^1)$, where $\underline{S}^1$ denotes smooth circle-valued functions.
\end{enumerate}
A key result in \v{C}ech cohomology is that for sheaves that admit partitions of unity, such as $C^k(X,\Omega^p)$, the \v{C}ech cohomology groups $\check{H}^k(X,\Omega^p)$ are trivial for $k>0$: every cocycle is a coboundary. 


\subsubsection*{Hermitian line bundles:  conventions}
Consider now a Hermitian line bundle $\pi : L \to X$ with transition functions $g_{\alpha\beta} : U_{\alpha\beta} \to S^1$ obeying 
\begin{align}
g_{\alpha\beta} g_{\beta\gamma} g_{\gamma\alpha} = 1~
\end{align} 
on all non-empty triple overlaps $U_{\alpha\beta\gamma}$, as well as connection $1$-forms $A_\alpha$ defined on each $U_{\alpha}$ and satisfying
\begin{align}
A_\alpha &  = A_\beta -i d \log g_{\alpha\beta}
\end{align}
on each $U_{\alpha\beta}$.   Gauge transformations are encoded by functions $h_\alpha : U_\alpha \to S^1$, which act by
\begin{align}
g_{\alpha\beta} &\to g_{\alpha\beta} h_\alpha (h_\beta)^{-1}~, &
A_{\alpha} & \to A_{\alpha} - i d\log h_\alpha~.
\end{align}
A gauge transformation is global if on every overlap $h_\alpha = h_\beta$, so that $h_\alpha$ is a restriction of a circle-valued function defined on $X$ to the set $U_\alpha$.

This data is elegantly presented in the \v{C}ech language: the data for a line bundle with connection is a pair $(g,A)$ with  $g \in C^1(X, \underline{S}^1)$,  $A \in C^0(X,\Omega^1)$ subject to
\begin{align}
\delta g & =1~, &
\delta A & = -id \log g~,
\end{align}
and gauge transformations are encoded by $h \in C^0(X, \underline{S}^1)$, and they act by
\begin{align}
g &\to g \delta h~, &
A &\to A -i d\log h~.
\end{align}
A global gauge transformation satisfies $\delta h = 1$.  

Two bundles $L\to X$, $L'\to X$ with data $(g,A)$ and $(g',A')$ respectively, are isomorphic if and only if it is possible (after a suitable refinement of the covers) to find a gauge transformation $h$ such that $(g',A') = (g \delta h, A-id\log h)$.

In what follows, we will take all of our line bundles to be Hermitian, so that the transition functions are circle-valued. 

\subsubsection*{Equivariant line bundles}
Given a line bundle $L \to X$ with transition functions $g$ and connection $A$, we would like to lift the action of $G$ on $X$ to an action on $L$.   For every $a \in G$ we can use $\varphi_a$ to construct the pull-back bundle $\varphi_a^\ast(L)$.  We say $L \to X$ is $G$-equivariant if and only if $\varphi_a^\ast(L) \simeq L$, and the isomorphisms, which we denote by $R^a \in C^0(X,\underline{S}^1)$ are compatible with the group multiplication law.  That is, we have the diagram
\begin{equation}
\begin{tikzcd}
\varphi_a^\ast(L) \ar[d] & \simeq & L  \ar[d] \\
X \ar[rr, "\varphi_a"] & ~& X
\end{tikzcd}
\end{equation} 
and for any section of $L$, $s \in C^0(X, \Omega^0)$, we have
\begin{align}
\varphi_a^\ast (s)= s(\varphi_a(x)) & = R^a(x) s(x)~.
\end{align}
Since $\varphi_{ab}^\ast(s) = \varphi_b^\ast \varphi_a^\ast(s)$,
\begin{align}
R^{ab} s & = R^b\varphi_b^\ast(R^a) s~,
\end{align}
for any section $s$, and we conclude that for all $a,b\in G$
\begin{align}
\label{eq:Rcomposition}
R^{ab} & = R^b \varphi_b^\ast(R^a)~.
\end{align}
Since $s$ also satisfies $\delta s = g$---this is the \v{C}ech language for the perhaps more familiar relation $s_\alpha = g_{\alpha\beta} s_b$ for all $U_{\alpha\beta}$---consistency with the pull-back requires
\begin{align}
\label{eq:gpullback}
\varphi_a^\ast(g) & = g \delta R^a~.
\end{align}
This implies that the $G$-action preserves the first Chern class of the bundle, i.e. $\varphi_a^\ast(c_1(L)) = c_1(L)$, which is a necessary condition for the lift to exist.  We will be interested in lifts that also preserve the connection, meaning that the covariant derivative $D s = ds - i As$ satisfies $\varphi_a^\ast (Ds) = R^a (Ds)$.  This requires
\begin{align}
\label{eq:Apullback}
\varphi_a^\ast(A) & = A -id \log R^a~.
\end{align}
Notice that the $G$--action preserves the curvature $F = dA$:  $\varphi_a^\ast(F) = F$.

It is not hard to check that the definition is consistent with gauge transformations:  if $R^a$ provide the lift for the bundle data $(g,A)$, then 
\begin{align}
R^a_{\text{new}} = R^a \frac{\varphi_a^\ast(f)}{f}
\end{align}
provide a lift for the gauge equivalent data $(g_{\text{new}},A_{\text{new}}) = (g \delta f, A - i d\log f)$.

Geometrically, all of this amounts to finding a $G$-action on the total space of the line bundle that is consistent with the projection and choice of connection.
%
%

\subsection{Gerbes: basic structure} \label{ss:appgerbes}
Having reviewed the case of line bundles, we now extend the discussion to $G$-equivariant gerbes.  We begin with the defining data of a gerbe over $X$ with connection: $(\vartheta,\beta, B)$, 
\begin{align}
\vartheta &\in C^2(X,\underline{S}^1)~,&
\beta &\in C^1(X, \Omega^1)~, &
B &\in C^0(X,\Omega^2)~&
\end{align}
satisfying
\begin{align}
\delta \vartheta &= 1~,&
\delta \beta & = i d\log\vartheta~,&
\delta B & = d \beta~.&
\end{align}

Two gerbes with connections are equivalent if and only if (after a suitable refinement of cover) they are related by a 0- and 1-gauge transformations, with parameters $f \in C^1(X,\underline{S^1})$ and $\eta \in C^0(X, \Omega^1)$, which act by
\begin{align}
\vartheta &\to \vartheta \delta f~, &
\beta &\to \beta +i d \log f + \delta \eta~, &
B & \to B + d\eta~.
\end{align}
We will call the full transformation $(f,\eta)$ a gerbe gauge transformation.   Note that $(f,\eta)$ leave the gerbe data invariant if and only if $f$ defines transition functions for a flat line bundle over $X$ with a compatible flat connection $\eta$.

The gerbe curvature $H = \ff{1}{2\pi} dB$ is a closed $3$-form on $X$.  Like the curvature of a line bundle, the curvature $H$ is gerbe gauge-invariant, and its cohomology class characterizes gerbes at the level of topology.

\subsubsection*{Vertical flat gerbes}
Given a closed $1$-form $\Theta$ on $X$ and a flat line bundle $L \to X$ with data $(g,\Lambda)$, we obtain a flat connection on a trivial gerbe by setting
\begin{align}
\vartheta & = 1~,&
\beta & = i \log(g) \Theta~,&
B & = \Lambda \wedge \Theta~.
\end{align}
If the transition functions are constant (such a choice is always possible for a flat line bundle), then $\beta$ can be gauged away while keeping $\vartheta =1$.  Of course this is exactly the situation when the connection $\Lambda$ is a global $1$-form on $X$, and these are gerbes with which we are concerned in the main text.

\subsection{$G$-action on the gerbe data}

In the situation of a line bundle, we had a clear geometric perspective on finding a lift of the $G$--action to $L$:  we were essentially finding a set of diffeomorphisms on the total space of the line bundle compatible with the projection to $X$.  Such a perspective is not immediately available for a gerbe, but we can, as in~\cite{Sharpe:2000ki}, study the action on $(\vartheta,\beta,B)$, and we will show that the solution to the consistency requirements of that action allows for more general results than those obtained in~\cite{Sharpe:2000ki}.

Starting with the defining relations of the gerbe we demand that for every $a\in G$ we have $(f^a,\eta^a)$ such that
\begin{align}
\label{eq:gerbesimplepullbacks}
\varphi_a^\ast(\vartheta) & = \vartheta \delta f^a~, &
\varphi_a^\ast(\beta) & = \beta + i d\log f^a + \delta \eta^a~, &
\varphi_a^\ast(B) & = B + d\eta^a~.
\end{align}
Applying $\varphi_b^\ast$, we find that consistency with the group law require
\begin{align}
\vartheta \delta f^{ab} & = \vartheta  \delta f^b \varphi_b^\ast(\delta f^a)~, \nonumber\\
- i d\log \frac{ f^{ab}}{f^b \varphi_b^\ast (f^a)} & = \delta \left(\eta^{ab} - \eta^b - \varphi_b^\ast(\eta^a) \right)~,\nonumber\\
0 & =d \left(\eta^{ab} - \eta^b - \varphi_b^\ast(\eta^a) \right)~.
\end{align}
These conditions are solved by
\begin{align}
\label{eq:generalgrouplaw}
f^{ab} & = f^b \varphi_b^\ast( f^a) k^{a,b}~,&
\eta^{ab} & =   \eta^b + \varphi_b^\ast(\eta^a)+ \tau^{a,b}~,
\end{align}
where $(k^{a,b},\tau^{a,b})$ is the data for a flat line bundle $T^{a,b} \to X$.

\subsubsection*{Associativity}
There are non-trivial conditions from associativity of the group product.  Imposing $f^{a(bc)} = f^{(ab)c}$ and $\eta^{a(bc)} = \eta^{(ab)c}$ in~(\ref{eq:generalgrouplaw}), requires
\begin{align}
\label{eq:TTisomorphismsbundles}
T^{a,bc} \otimes T^{b,c}&\simeq T^{ab,c} \otimes \varphi_c^\ast (T^{a,b} ) ~,
\end{align}
with
\begin{align}
\label{eq:TTisomorphisms}
k^{a,bc} k^{b,c} &= k^{ab,c} \varphi_c^\ast(k^{a,b})~,
&
\tau^{a,bc} +\tau^{b,c} &= \tau^{ab,c}+ \varphi_c^\ast(\tau^{a,b})~.
\end{align}


Any flat connection $\tau^{a,b}$ can be written as
\begin{align}
\tau^{a,b} =  i d\log h^{a,b}~
\end{align}
for some $h^{a,b} \in C^0(X, \underline{S}^1)$ satisfying
\begin{align}
\delta h^{a,b} = t^{a,b}  \left(k^{a,b}\right)^{-1}~
\end{align}
for some \v{C}ech cocycle $t^{a,b} \in C^1(X,S^1)$.  Recall that any flat line bundle admits a trivialization with constant transition functions, and the choice of such constant transition functions is a flat structure on the bundle~\cite{Kobayashi:1987VB};  $t^{a,b}$ is such a flat structure on the line bundle $T^{a,b} \to X$.

The relations~(\ref{eq:TTisomorphisms}) are equivalent to
\begin{align}
\label{eq:associativity}
\delta \left( \frac{ h^{ab,c} \varphi_c^\ast(h^{a,b})}{h^{a,bc} h^{b,c} } \right) & = \frac{t^{ab,c} \varphi_c^\ast(t^{a,b})}{t^{a,bc} t^{b,c}}~,
&
d \left( \frac{ h^{ab,c} \varphi_c^\ast(h^{a,b})}{h^{a,bc} h^{b,c} } \right) & = 0~.
\end{align}
Therefore for every $a,b,c\in G$ there exists a locally constant phase $\psi^{a,b;c} \in C^0(X,S^1)$ such that
\begin{align}
\label{eq:Psimagic}
\varphi_c^\ast(h^{a,b}) &= \psi^{a,b;c}\frac{ h^{a,bc} h^{b,c}}{h^{ab,c}}~, &
t^{a,bc} t^{b,c} & = \delta (\psi^{a,b;c}) t^{ab,c} \varphi_c^\ast(t^{a,b})~.
\end{align}
So, we see that $\psi^{a,b;c}$ is a constant gauge transformation relating the flat structures on the flat bundles in~(\ref{eq:TTisomorphismsbundles}).  Letting $\varphi_d^\ast$ act on both sides of the first equation in~(\ref{eq:Psimagic}), we find that $\psi$ is a group cocycle, i.e. for each patch $U_\alpha$, the constants $\psi^{a,b;c}_\alpha$ define a class in the group cohomology $H^3(G,\GU(1))$.

The appearance of the line bundles $T^{a,b}$ is a new feature compared to the analysis of~\cite{Sharpe:2000ki}, and it in principle allows for more general $G$-actions.  As with many matters involving non-trivial gerbe structures, it would be useful to have concrete classes of examples that realize these seemingly more exotic possibilities.

\subsubsection*{Discrete torsion}
From the first of the expressions in~(\ref{eq:Psimagic}) we can already see the possibility of discrete torsion of~\cite{Vafa:1986wx}.  Let $\lambda^{a,b}$ be a $G$ 2-cocycle, i.e.  $\lambda:  G\times G \to \GU(1)$, such that
\begin{align}
\frac{\lambda^{a,bc} \lambda^{b,c}}{\lambda^{ab,c} \lambda^{a,b}} = 1~.
\end{align}
In that case, if we have found $h^{a,b}$ that satisfy~(\ref{eq:Psimagic}) with some $\psi^{a,b;c}$, then we obtain a new solution to~(\ref{eq:Psimagic}) with the same $\psi^{a,b;c}$ by setting
\begin{align}
h_{\text{new}}^{a,b} = h^{a,b} \lambda^{a,b}~.
\end{align}
As we will see below, the factors $h^{a,b}$ only enter the orbifold CFT partition function through combinations $h^{a,b}/h^{b,a}$ for commuting elements $a,b \in G$.  Therefore, shifting $\lambda^{a,b} \to \lambda^{a,b} \xi^{a} \xi^{b} /\xi^{ab}$ for any $G$ 1-cochain $\xi$ leave the partition function invariant; said another way, the partition function only depends on the cohomology class $[\lambda] \in H^2(G,\GU(1))$.

\subsubsection*{Action on a topologically trivial gerbe}
As for line bundles, the familiar exponential short exact sequence
\begin{equation}
\label{eq:ExpSES}
\begin{tikzcd}
0 \ar[r] & \Z \ar[r] & \R \ar[r,"e^{2\pi i \cdot}"] & \GU(1) \ar[r] & 0~
\end{tikzcd}
\end{equation}
and the associated long exact sequence in cohomology can be used to relate differential data encoded in the connections to topological information.  For example, the cohomology class of the curvature $[H] \in H^3(X,\R)$ of a gerbe is the image of a  class in $H^3(X,\Z)$, and this latter class characterizes the gerbe at the level of topology.

A gerbe is flat if and only if its curvature vanishes: $H = 0$.   Flat gerbes are classified by the cohomology group $H^2(X, \GU(1))$~\cite{Hitchin:1999fh}, which encodes the holonomy of the $B$-field on $2$-cycles.  Using the long exact sequence associated to~(\ref{eq:ExpSES}), we find
\begin{equation}
\label{eq:flatgerbecharacterization}
\begin{tikzcd}
0 \ar[r] & H^2(X,\R)/H^2(X,\Z) \ar[r] & H^2(X,\GU(1)) \ar[r] & \left\{ H^3(X,\Z)\right\}^{\text{tors}} \ar[r] & 0~,
\end{tikzcd}
\end{equation}
where the last term is the torsion subgroup of $H^3(X,\Z)$.   The right-hand-side encodes a possible non-trivial topology of the flat gerbe, while the left-hand-side encodes the choice of $B$ up to gauge transformations. This is analogous to the characterization of connections on flat line bundles, where the same description holds with cohomology degrees reduced by $1$.

A gerbe with data $(\vartheta,\beta,B)$ is topologically trivial if and only if $\vartheta$ is a \v{C}ech coboundary.  For any topologically trivial gerbe it is possible to make a gerbe gauge transformation that sets $\vartheta = 1$ and $\beta = 0$.  This a partial gauge fixing, and the gerbe gauge transformations that preserve the choice are $(f,\eta) = (g,\cA)$, where $(g,\cA)$ is data for a line bundle $\cL \to X$; these act by
\begin{align}
B \to B + d\cA~.
\end{align}

We now consider the $G$-action on a topologically trivial gerbe.  Following our previous line of reasoning, we specify line bundles $\cL^a \to X$ with data $(g^a,\cA^a)$ for every $a\in G$ such that
\begin{align}
\label{eq:bundleactionB}
\varphi_a^\ast(B) & = B + d\cA^{a}~.
\end{align}
The group structure imposes further requirements on this data.  Specializing~(\ref{eq:generalgrouplaw}) to the topologically trivial gerbe, we find that the data for the flat $T^{a,b}$ bundles is determined by the data for the $\cL^a$ bundles that specify the $G$-action.  Denoting the dual bundle to $L$ by $L^\vee$, 
\begin{align}
\label{eq:gAconditions1}
T^{a,b} &\simeq \cL^{ab} \otimes \left(\cL^b\otimes\varphi_b^\ast(\cL^a)\right)^{\vee}~,&
 t^{a,b} &=
 \frac{g^{ab}}{g^b \varphi_b^\ast (g^a)} \delta h^{a,b}~,&
 i d\log h^{a,b} & = \cA^{ab} - \cA^b-\varphi_b^\ast (\cA^a)~.
\end{align}

\subsubsection*{Bundle and gerbe gauge transformations}
Are these structures are consistent with gauge transformations of the gerbe and of the individual line bundles $\cL^a$?  If the latter does not hold, then we could not speak of the $G$-action as encoded in a choice of line bundles; if the former does not hold, then our $G$-action would depend on a particular representative $B$.  We dispel both of these concerns, starting with bundle gauge transformations.

Given $(g^a,\cA^a)$ and $h^{a,b}$ satisfying~(\ref{eq:Psimagic}),~(\ref{eq:bundleactionB}), and~(\ref{eq:gAconditions1}), we can pick gauge-equivalent data for the $\cL^a$
\begin{align}
\label{eq:bundletransforms}
g_{\text{new}}^a &= g^a f^a~, &
\cA_{\text{new}}^a & = \cA^a - i d\log f^a~,
\end{align}
and set
\begin{align}
h^{a,b}_{\text{new}} =  h^{a,b}  f^b \varphi_b^\ast(f^a){f^{ab}}^{-1}~.
\end{align}
It is straightforward to check that the factor $\xi^{a,b} =  f^b \varphi_b^\ast(f^a)(f^{ab})^{-1}$---a gauge transformation on $T^{a,b}$---satisfies
\begin{align}
\frac{ \xi^{ab,c} \varphi_c^\ast(\xi^{a,b})}{\xi^{a,bc} \xi^{b,c} }  = 1~,
\end{align}
So, we see that $(g^a_{\text{new}},\cA^a_{\text{new}})$ give a consistent $G$-action.

Next, consider a gerbe gauge transformation specified by a line bundle $L \to X$ with data $(g_L, \Lambda)$ and
\begin{align}
B_{\text{new}} = B + d \Lambda~.
\end{align}
Suppose that we have a $G$-action on the gerbe with connection $B$ specified by line bundles $\cL^a$ such that
\begin{align}
\varphi_a^\ast(B) = B + d\cA^a~.
\end{align}
We can then obtain a $G$-action on the gerbe with connection $B_{\text{new}}$ by taking new bundles
\begin{align}
\cL^a_{\text{new}} &= \cL^a \otimes \varphi_a^\ast(L) \otimes L^\vee~,
\end{align}
with data accordingly satisfying
\begin{align}
\label{eq:gerbegaugeA}
g^a_{\text{new}} &= g^a \varphi_a^\ast(g_L) (g_L)^{-1}~, &
\cA^a_{\text{new}} & = \cA^a  + \varphi_a^\ast(\Lambda) - \Lambda~.
\end{align}

\subsubsection*{Trivial $T^{a,b}$}
In the special situation that $t^{a,b} =1$, ~(\ref{eq:gAconditions1}) gives isomorphisms
\begin{align}
\label{eq:bundleproduct}
\cL^b\otimes\varphi_b^\ast(\cL^a) &\simeq \cL^{ab} ~,&
g^b \varphi_b^\ast(g^a)  & = g^{ab}\delta h^{a,b}~,&
\cA^b + \varphi_b^\ast(\cA^a)  & = \cA^{ab}- i d\log h^{a,b}~.
\end{align}
As we discussed in section~\ref{ss:cechlinebundle}, such isomorphisms have a geometric interpretation at the level of sections, or, equivalently, as diffeomorphisms on the total space of the bundle that commute with projection to $X$:  for a point $(x,\xi)$ in the total space of $\cL^b\otimes\varphi_b^\ast(\cL^a)$ we have
\begin{align}
 \cL^b\otimes\varphi_b^\ast(\cL^a)&\to\cL^{ab} ~,&
(x,\xi) &\mapsto (x, h^{a,b}(x)\xi )~.
\end{align}
The associativity of the group multiplication law then requires the diffeomorphisms to satisfy
\begin{align}
\label{eq:Rab}
\varphi_c^\ast(h^{a,b} ) h^{ab,c} = h^{a,bc} h^{b,c}~.
\end{align}
Using these relations in~(\ref{eq:Psimagic}), we see that $\psi^{a,b;c} =1$.   A bit more generally, if $t^{a,b} = \delta s^{a,b}$ for some $s^{a,b} \in C^0(X,S^1)$, then we can absorb the $s^{a,b}$ into a redefinition of $h^{a,b}$ while preserving~(\ref{eq:bundleproduct}).  The associativity of the bundle isomorphisms then again implies~(\ref{eq:Rab}), so that the $\psi^{a,b;c} = 1$.  We also see that (\ref{eq:Rab}) is consistent with multiplying $h^{a,b}$ by a representative of $H^2(G,\GU(1))$, and modifying it by a group coboundary can be absorbed into the gauge transformations of the individual bundles $\cL^a$---again, this is the discrete torsion of~\cite{Vafa:1986wx}.  In this way, we recover the results of~\cite{Sharpe:2000ki} when $T^{a,b}$ are trivial bundles.

\subsection{A trivial flat gerbe on $\R^d$ and orbifold CFT}
Following~\cite{Sharpe:2000ki}, we now discuss how the equivariant structure just defined allows us to define the contribution of the $B$-field to the orbifold partition function for the theory on $X/G$.  This is not easy for a general $X$ equipped with a gerbe, but it is manageable and instructive in the special case that $X = \R^d$, $G$ is abelian, and the gerbe is trivial and flat.\footnote{The results are also relevant for non-abelian $G$, since the orbifold construction restricts the $a,b$ to mutually commuting elements in $G$.}  

Taking the worldsheet to be a torus, we fix a map from $T^2 \to X$ with image $S_{a,b}(x)$:
\begin{equation}
\begin{tikzpicture}
\draw[thin] (0,0) node {$\bullet$} node [below] {$x$} -- (2,0) node {$\bullet$} node [below] {$\varphi_b(x)$}  -- (3,1) node{$\bullet$}node[right] {$\varphi_{ab}(x)$} -- (1,1) node {$\bullet$} node[above] {$\varphi_a(x)$} --cycle;
\end{tikzpicture}
\end{equation}
We then define the phase factor
\begin{align}
\label{eq:bigphase}
P_{a,b} = \exp i\left[ \int_{S_{a,b}(x)} B\right] \exp i \left[ \int_x^{\varphi_a(x)} \cA^b - \int_x^{\varphi_b(x)} \cA^a\right] 
\frac{ h^{a,b}(x)}{h^{b,a}(x)}~.
\end{align}
This phase factor is precisely the holonomy of the gerbe on $X/G$ associated to the cycle that lifts to $S_{a,b}(x)$, and it enjoys three important properties.
\begin{enumerate}
\item It is independent of the basepoint $x$.

First observe that $\int_{S_{a,b}(x)} B$ is base-point independent because $S_{a,b}(x)$ and $S_{a,b} (x+v)$ are homologous for any $v \in \R^d$ and $H=0$.  So, under a variation of $x\to x+ v$, we have
\begin{align}
-i \delta_v \log P_{a,b} &= \text{Lie}_v \left[ \int_x^{\varphi_a(x)} \cA^b - \int_x^{\varphi_b(x)} \cA^a\right]  - i \text{Lie}_v \log \frac{h^{a,b}}{h^{b,a}}(x)~ \nonumber\\
& = v \llcorner \left( \varphi_a^\ast(\cA^b) - \cA^b - \varphi_b^\ast(\cA^a) + \cA^a\right) - i v \llcorner d \log \frac{h^{a,b}}{h^{b,a}} = 0~.
\end{align}

\item It is invariant under gerbe gauge transformations.

This follows because
\begin{align}
\int_{S_{a,b}} B_{\text{new}} &=  \int_{S_{a,b}} B 
+\left( \int_x^{\varphi_a(x)} - \int_x^{\varphi_b(x)} 
          +\int_{\varphi_a(x)}^{\varphi_{ab}(x)}
          - \int_{\varphi_b(x)}^{\varphi_{ab}(x)}\right)  \Lambda \nonumber\\
& =\int_{S_{a,b}} B 
+ \int_x^{\varphi_a(x)} \left( \Lambda - \varphi_b^\ast(\Lambda)\right)
- \int_x^{\varphi_b(x)} \left(\Lambda - \varphi_a^\ast(\Lambda) \right)~,
\end{align}
and the last two terms are canceled by the $\Lambda$ transformations of the $\cA^a$ from~(\ref{eq:gerbegaugeA}).

\item It is invariant under bundle gauge transformations.

To see this, we note that $B$ does not transform under~(\ref{eq:bundletransforms}), while
\begin{align}
\left[ \int_x^{\varphi_a(x)} \cA^b - \int_x^{\varphi_b(x)} \cA^a\right] \to
\left[ \int_x^{\varphi_a(x)} \cA^b - \int_x^{\varphi_b(x)} \cA^a\right]  \times \frac{ f^a\varphi_a^\ast(f^b)}{f^b \varphi_b^\ast(f^a)}~.
\end{align}
The last factor is then canceled by the transformation of the ratio $h^{a,b}/h^{b,a}$.

\end{enumerate}
While these properties make the factor well-defined, it is not obvious that including such a factor leads to a well-behaved orbifold CFT.  Fortunately, if we assume that the $X/G$ orbifold CFT is well-behaved without this phase factor, then the necessary and sufficient conditions for a well-defined partition function at any genus are well-known from the classic work~\cite{Vafa:1986wx}:  the phases should satisfy
\begin{align}
P_{a,b} P_{b,a} & = 1~, &
P_{a,a} & = 1~, &
P_{ab,c} &= P_{a,c} P_{b,c}~.
\end{align}
Using~(\ref{eq:bigphase}) we see that
\begin{align}
P_{a,b} P_{b,a} = \exp i \left[ \int_{S_{a,b}(x)} B + \int_{S_{b,a}(x)} B \right]~.
\end{align}
The two integrals cancel because $S_{a,b}$ and $-S_{b,a}$ are homologous, and $H=0$.  Moreover, since the integrals cancel exactly, we also see that $P_{a,a} = 1$.

To study the last condition, we write
\begin{align}
P_{a,b} = P_{a,b}^B P_{a,b}^{\cA} P_{a,b}^{h}~
\end{align}
and study the ratios for the different terms in turn.  

The ratio of the $B$- factors,
\begin{align}
\frac{ P_{ab,c}^B }{ P^B_{a,c} P^B_{b,c} }~,
\end{align}
involves the integral $\int_{S_{ab,c}} B$.  We can relate this to the other integrals by constructing a closed surface as follows (we drop the $x$ dependence and just label the points by elements of $G$):
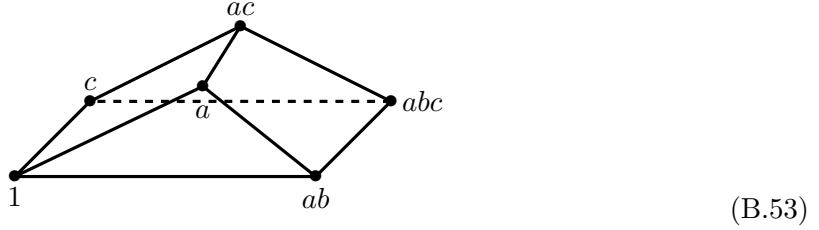
\begin{equation}
\label{eq:rooftop}
\begin{tikzpicture}
\draw
(0,0) node {$\bullet$} node [below] {$1$} -- 
(4,0) node {$\bullet$} node [below] {$ab$}  -- 
(5,1) node{$\bullet$}node[right] {$abc$};
\draw[dashed] (5,1) --(1,1) node {$\bullet$} node[above] {$c$} ;
\draw (1,1)-- (0,0);
\draw (2.5,1.2) node {$\bullet$} node[below=3pt] {$a$};
\draw (3,2) node {$\bullet$} node[above] {$ac$};
\draw (0,0) -- (2.5,1.2) -- (4,0);
\draw (2.5,1.2) -- (3,2) -- (5,1);
\draw (3,2) -- (1,1);
\end{tikzpicture}
\end{equation}
Because $H =0$,
\begin{align}
\int_{S_{ab,c}(x)} B = \int_{S_{a,c}(x)} B + \int_{S_{b,c}(\varphi_a(x))} B-\int_{\Sigma_{a,b}(x)} B + \int_{\Sigma_{a,b}(\varphi_c(x))} B~,
\end{align}
where $\Sigma_{a,b}$ is the oriented surface with ordered vertices $1,a,ab$.  We then obtain
\begin{align}
\int_{S_{ab,c}(x)} B -  \int_{S_{a,c}(x)} B- \int_{S_{b,c}(x)} B &=
\int_{S_{b,c}(x)} (\varphi_a^\ast(B) - B)
+
\int_{\Sigma_{a,b}} (\varphi_c^\ast(B) - B) \nonumber\\
& = \int_{S_{b,c}(x)} d\cA^a + \int_{\Sigma_{a,b}} d\cA^c~.
\end{align}
So, we conclude
\begin{align}
\frac{ P_{ab,c}^B }{ P^B_{a,c} P^B_{b,c} } = 
\exp i \left[ \int_{S_{b,c}(x)} d\cA^a + \int_{\Sigma_{a,b}} d\cA^c\right]~.
\end{align}
Next we tackle the $P_{ab,c}^{\cA}$ factor.  Using~(\ref{eq:gAconditions1}) to eliminate the $\cA^{ab}$ term in $P^{\cA}_{ab,c}$ we arrive at
\begin{align}
\frac{P^{\cA}_{ab,c}}{P^{\cA}_{a,c}P^{\cA}_{b,c}} = 
\exp i \left[ \left( \int_x^{\varphi_{ab(x)} }- \int_x^{\varphi_a(x)} - \int_x^{\varphi_b(x)} \right) \cA^c
+ \int_x^{\varphi_c(x)} (\cA^a-\varphi_b^\ast(\cA^a) )
\right]\frac{\varphi_c^\ast(h^{a,b})}{h^{a,b}}~.
\end{align}
We rewrite the first term as
\begin{align}
\left( \int_x^{\varphi_{ab(x)} }- \int_x^{\varphi_a(x)} - \int_x^{\varphi_b(x)} \right) \cA^c = -\int_{\Sigma_{a,b}} d\cA^c~ -\int_x^{\varphi_b(x)} (\cA^c - \varphi_a^\ast(\cA^c) ),
\end{align}
Because $G$ is abelian, we have the relation
\begin{align}
\cA^a -\varphi_c^\ast(\cA^a) +id\log h^{c,a} = 
\cA^c -\varphi_a^\ast(\cA^c) +id\log h^{a,c}~,
\end{align}
so that 
\begin{align}
\left( \int_x^{\varphi_{ab(x)} }- \int_x^{\varphi_a(x)} - \int_x^{\varphi_b(x)} \right) \cA^c = -\int_{\Sigma_{a,b}} d\cA^c~ -\int_x^{\varphi_b(x)} (\cA^a - \varphi_c^\ast(\cA^a) 
+ id \log \ff{h^{c,a}}{h^{a,c}} )~,
\end{align}
and combining factors we learn that
\begin{align}
\frac{ P_{ab,c}^B }{ P^B_{a,c} P^B_{b,c} }\frac{P^{\cA}_{ab,c}}{P^{\cA}_{a,c}P^{\cA}_{b,c}} 
&= \exp i \left[\int_{S_{b,c}(x)} d\cA^a + \int_x^{\varphi_c(x)} (\cA^a-\varphi_b^\ast(\cA^a)-\int_x^{\varphi_b(x)} (\cA^a - \varphi_c^\ast(\cA^a)\right] \nonumber\\
&\qquad\times \frac{\varphi_c^\ast(h^{a,b})}{h^{a,b}}\varphi_b^{\ast}\left(\frac{h^{c,a}}{h^{a,c}}\right) \frac{h^{a,c}}{h^{c,a}}
\end{align}
The square bracket is zero, and we have
\begin{align}
\frac{ P_{ab,c}^B }{ P^B_{a,c} P^B_{b,c} }\frac{P^{\cA}_{ab,c}}{P^{\cA}_{a,c}P^{\cA}_{b,c}} =\frac{\varphi_c^\ast(h^{a,b})}{h^{a,b}}\varphi_b^{\ast}\left(\frac{h^{c,a}}{h^{a,c}}\right) \frac{h^{a,c}}{h^{c,a}}~.
\end{align}
Combining this with the ratio $P^{h}_{ab,c} / P^h_{a,c} P^h_{b,c}$, we find
\begin{align}
\label{eq:almostthere}
\frac{P_{ab,c}}{P_{a,c} P_{b,c}} = 
\frac{h^{ab,c}}{h^{c,ab}} \frac{h^{c,b}}{h^{b,c}}\frac{\varphi_c^\ast(h^{a,b})}{h^{a,b}}\varphi_b^{\ast}\left(\frac{h^{c,a}}{h^{a,c}}\right) ~.
\end{align}
Using the constraint from~(\ref{eq:Psimagic}) we can eliminate the pullbacks from this expression, and we find the necessary and sufficient condition for a well-defined CFT partition function at any genus:
\begin{align}
\label{eq:finalphase}
\frac{P_{ab,c}}{P_{a,c} P_{b,c}} = \frac{ \psi^{a,b;c} \psi^{c,a;b}}{\psi^{a,c;b}} = 1~.
\end{align}
We saw above that in the case $t^{a,b} = 1$ we have $\psi^{a,b;c} = 1$, so the orbifold partition function is consistent for every equivariant gerbe with trivial $T^{a,b}$.

\subsubsection*{Possibilities for new equivariant gerbes}

While trivial $T^{a,b}$ imply that $\psi^{a,b;c}=1$, the consistency condition in~(\ref{eq:finalphase}) appears to be weaker than $\psi^{a,b;c} = 1$.  Since the general requirements for a $G$--action on the gerbe--- (\ref{eq:Psimagic}) for a general gerbe or~(\ref{eq:bundletransforms}) for a trivial gerbe---also do not appear to require $\psi^{a,b;c}=1$, this naively suggests that there may be consistent equivariant gerbes with $\psi^{a,b;c} \neq 1$ satisfying~(\ref{eq:finalphase}).  We do not believe this to be the case:  for example, it is hard to understand how a non-trivial class in $H^3(G,\GU(1))$ would show up in the computation of $H^2(X/G, \GU(1))$---the cohomology group that classifies flat gerbes with connections.

On the other hand, it seems more likely that it is possible to include non-trivial bundles $T^{a,b}$, provided that~(\ref{eq:TTisomorphismsbundles}) is obeyed with $\psi^{a,b;c} =1$.   This could potentially enlarge the class of equivariant gerbes on $X$ with $\left(H^2(X,\Z)\right)^{\text{tors}} \neq 0$.    If our goal is to use the equivariant gerbes to describe gerbes on the quotient $M = X/G$, then we can avoid the complications associated to the $T^{a,b}$ by replacing $X$ with its universal cover $X^\circ$, so that $X = X^\circ/G^\circ$.  Since $(H^2(X^\circ,\Z))^{\text{tors}} = (H_1(X^\circ,\Z))^{\text{tors}} = 0$, all flat line bundles on $X^\circ$ are trivial, and we can describe the gerbes on $M$ with the simpler construction of equivariant gerbes on $X^\circ$ with respect to the group $G^\circ \times G$.  

\section{$X_{k,\vec{k}}$ for $\Z_{N>2}$}
In this appendix we describe the spaces obtained for each $\Ttor^5/\Z_N$ quotient with $N=3,4,6$.  These results match the $\Z_N$ quotients described in section 4 of~\cite{deBoer:2001wca}, as well as the classification results obtained in~\cite{Fischer:2012qj}.

\subsection*{$\Z_3$ quotients}
The action on $\Ttor^4$ is
\begin{align}
\varphi_a (\vec{x}) & = \begin{pmatrix} -1 & 1 & 0 & 0 \\ -1 & 0 &0 & 0 \\ 0 & 0 &-1 &1 \\ 0 & 0 & -1& 0 \end{pmatrix} \vec{x}~.
\end{align}
Focusing on a single $\Ttor^2$ factor, we see that 
\begin{align}
M = \varphi_a^T - \iden = \begin{pmatrix} -2 & -1 \\ 1 & -1 \end{pmatrix}~
\end{align}
generates an index $3$ sublattice in $\Z^2$, which allows us to set $k_1 =0$, while
letting $k_2 \in \{0,1,2\}$.  Examining the quotients in detail, we find that each quotient is diffeomorphic to one of the following three types:
\begin{align*}
\text{orbifold} && \text{quotient action} \\[2mm]
X_{0,0} = \Ttor^4/\Z_3 \times S^1 &&  (\varphi_a(\vec{x}), x_5) \\
X_{1,0} ~\text{(smooth)}&&  (\varphi_a(\vec{x}), x_5+\ff{1}{3}) \\
X_{0,\vec{k}}~\text{(singular)} &&  (\varphi_a(\vec{x}), x_5+x_3)
\end{align*}

\subsection*{$\Z_4$ quotients}
The action on $\Ttor^4$ is
\begin{align}
\varphi_a (\vec{x}) & = \begin{pmatrix} -\ep & 0 \\ 0 &-\ep \end{pmatrix} \vec{x}~, &
\ep &= \begin{pmatrix} 0 & 1 \\-1 & 0 \end{pmatrix}~.
\end{align}
Focusing on a single $\Ttor^2$ factor we have
\begin{align}
M = \varphi_a^T - \iden = \begin{pmatrix} -1 & 1 \\ -1 & -1 \end{pmatrix}~.
\end{align}
This generates an index $2$ sub-lattice and allows us to set $k_2=0$ and $k_1 \in \{0,1\}$, and similarly for the second $\Ttor^2$ factor.  The diffeomorphism classes are then represented by 
\begin{align*}
\text{orbifold} && \text{quotient action} \\[2mm]
X_{0,0} = \Ttor^4/\Z_N \times S^1 &&  (\varphi_a(\vec{x}), x_5) \\
X_{1,0}~\text{(smooth)} &&  (\varphi_a(\vec{x}), x_5+\ff{1}{4}) \\
X_{2,0} &&  (\varphi_a(\vec{x}), x_5+\ff{1}{2})  \\
X_{0,\vec{k}} &&  (\varphi_a(\vec{x}), x_5+x_1)  
\end{align*} 

\subsection*{$\Z_6$ quotients}
The action on $\Ttor^4$ is
\begin{align}
\varphi_a (\vec{x}) & = \begin{pmatrix} 1 & -1 & 0 & 0 \\ 1 & 0 &0 & 0 \\ 0 & 0 &1 &-1 \\ 0 & 0 & 1& 0 \end{pmatrix} \vec{x}~,
\end{align}
For a single $\Ttor^2$ we find
\begin{align}
M = \varphi_a^T - \iden = \begin{pmatrix} 0& -1 \\ 1 & -1 \end{pmatrix}~,
\end{align}
generates the entire lattice, so that we can set $\vec{k} = 0$.  
  
There are then $4$ diffeomorphism types, represented by 
\begin{align}
\text{orbifold} && \text{quotient action} \\[2mm]
X_{k, 0} &&  (\varphi_a(\vec{x}), x_5+\ff{k}{6})~.
\end{align}

\bibliographystyle{./utphys}
\bibliography{./newref}

\end{document}